\documentclass[
aps,prx,twocolumn,english,superscriptaddress,
]{revtex4-2}

\usepackage{natbib}
\usepackage{amsfonts, amsmath, amssymb, mathrsfs, mathtools, amsthm, bbm, bm}
\usepackage{braket}
\usepackage{float}
\usepackage{afterpage}
\usepackage[version=4]{mhchem}
\usepackage[
    labelfont=bf, 
    font=small,
    justification=justified,
    format=plain
    ]{caption}
\usepackage[font=footnotesize,justification=justified]{subcaption}
\usepackage{comment}
\usepackage{tikz}
\usetikzlibrary{shadows,arrows.meta,positioning,backgrounds,fit}
\usepackage{quantikz}
\usepackage{booktabs, tabularx, colortbl}
\usepackage[linesnumbered,ruled]{algorithm2e}

\makeatletter
\def\BState{\State\hskip-\ALG@thistlm}
\makeatother

\newcolumntype{Y}{>{\raggedleft\arraybackslash}X}
\newcommand{\hl}[1]{#1} 

\theoremstyle{definition}

\DeclareMathOperator*{\argmax}{arg\,max}
\DeclareMathOperator*{\argmin}{arg\,min}
\DeclareMathOperator*{\argsort}{arg\,sort}
\newcommand{\angstrom}{\textup{\AA}}

\definecolor{plasma0}{rgb}{0.050383, 0.029803, 0.527975}
\definecolor{plasma1}{rgb}{0.417642, 0.000564, 0.658390}
\definecolor{plasma2}{rgb}{0.69284,  0.165141, 0.564522}
\definecolor{plasma3}{rgb}{0.881443, 0.392529, 0.383229}
\definecolor{plasma4}{rgb}{0.98826,  0.652325, 0.211364}

\makeatletter
\renewcommand*\env@matrix[1][*\c@MaxMatrixCols c]{%
  \hskip -\arraycolsep
  \let\@ifnextchar\new@ifnextchar
  \array{#1}}
\makeatother

\tikzset{%
  materia/.style={draw, fill=pink!30, text width=6.0em, text centered, minimum height=1.5em,drop shadow},
  etape/.style={materia, text width=11em, minimum width=10em, minimum height=3em, rounded corners, drop shadow},
  texto/.style={above, text width=6em, text centered},
  linepart/.style={draw, thick, color=black!50, -LaTeX, dashed},
  line/.style={draw, very thick, dotted, color=black!100, -LaTeX},
  ur/.style={draw, text centered, minimum height=0.01em},
  back group/.style={fill=yellow!10,rounded corners, draw=black!50, dashed, inner xsep=15pt, inner ysep=10pt},
}

\usepackage[bookmarks=false]{hyperref}

\begin{document}

\title{Contextual Subspace Variational Quantum Eigensolver Calculation of the Dissociation Curve of Molecular Nitrogen on a Superconducting Quantum Computer}

\author{Tim Weaving}
\affiliation{Centre for Computational Science, Department of Chemistry, University College London, WC1H 0AJ, United Kingdom}
\author{Alexis Ralli}
\affiliation{Centre for Computational Science, Department of Chemistry, University College London, WC1H 0AJ, United Kingdom}
\affiliation{Department of Physics and Astronomy, Tufts University, Medford, MA 02155, USA}
\author{Peter J. Love}
\affiliation{Department of Physics and Astronomy, Tufts University, Medford, MA 02155, USA}
\affiliation{Computational Science Initiative, Brookhaven National Laboratory, Upton, NY 11973, USA}
\author{Sauro Succi}
\affiliation{Center for Life Nano-Neuro Science @ La Sapienza, Italian Institute of Technology, 00161 Roma, Italy}
\affiliation{Department of Mechanical Engineering, University College London, WC1E 7JE, United Kingdom}
\affiliation{Department of Physics, Harvard University, Cambridge, MA 02138, USA}
\author{Peter V. Coveney}
\affiliation{Centre for Computational Science, Department of Chemistry, University College London, WC1H 0AJ, United Kingdom}
\affiliation{Advanced Research Computing Centre, University College London, WC1H 0AJ, United Kingdom}
\affiliation{Informatics Institute, University of Amsterdam, Amsterdam, 1098 XH, Netherlands}

\date{\today}

\begin{abstract}
In this work we present an experimental demonstration of the Contextual Subspace Variational Quantum Eigensolver on superconducting quantum hardware. In particular, we compute the potential energy curve for molecular nitrogen, where a dominance of static correlation in the dissociation limit proves challenging for many conventional quantum chemistry techniques. Our quantum simulations retain good agreement with the full configuration interaction energy in the chosen STO-3G basis, outperforming \hl{all benchmarked single-reference wavefunction techniques in capturing the bond-breaking appropriately. Moreover, our methodology is competitive with several multiconfigurational approaches, but at a considerable saving of quantum resource, meaning larger active spaces can be treated for a fixed qubit allowance.} To achieve this result we deploy an error mitigation/suppression strategy comprised of dynamical decoupling, measurement-error mitigation and zero-noise extrapolation, in addition to circuit parallelization that not only provides passive averaging of noise but improves the effective shot-yield to reduce the measurement overhead. Furthermore, we introduce a modification to previous adaptive ansatz construction algorithms that incorporates hardware-awareness \hl{into our variational circuits to minimize the transpilation cost for the target qubit topology.}
\end{abstract}

\maketitle

\section{Introduction}\label{sec:intro}
Quantum chemistry has been investigated as an application of quantum computing for almost two decades~\cite{aspuru2005simulated}. Instances of the electronic structure problem are intrinsically quantum mechanical, are challenging for classical methods at the few-hundred qubit scale and are of scientific and commercial importance.  Considerable development of quantum algorithms for quantum chemistry has taken place. Progress in hardware has lead to many small demonstrations of quantum chemical calculations on noisy intermediate-scale quantum (NISQ) devices. These demonstrations evaluate both the practicality of quantum algorithms when deployed on real hardware and also evaluate NISQ hardware against a real-world application benchmark. \hl{Variational quantum algorithms have been studied extensively for their shallow circuits, making them appealing for NISQ applications where modest coherence times limit the depth of circuits that can be executed successfully.} The current state-of-the-art, summarized by a representative sample of variational quantum eigensolver (VQE) realizations in Table~\ref{tab:VQE_exp_todate}, show that much progress is required before quantum computers can challenge their classical counterparts. 

The goal of quantum computing for quantum chemistry is to achieve quantum advantage. This implies the existence of problems where all classical heuristics fail to produce adequate results, while a quantum algorithm succeeds in rendering its chemical features to sufficiently high accuracy to be scientifically useful. This means larger systems and/or basis sets than are classically tractable, although candidate advantage applications are not easy to find, as recently brought to attention~\cite{lee2023evaluating, chan2024quantum}. The threshold of ``chemical accuracy" is used as a typical indication of success which, in NISQ demonstrations, is taken to mean the quantum calculation achieves an absolute error below $43$~meV with respect to the numerically-exact result in the chosen basis. This is an abuse of terminology as small basis set calculations will not typically be chemically-accurate when compared against experiment, \hl{which is the ultimate arbiter of computational utility; it has been suggested that \textit{algorithmic} accuracy is more appropriate terminology \cite{motta2022emerging}.} The point of interest is whether NISQ devices \hl{are able to achieve sufficient resolution such that, with realistic basis sets, we may address chemically-relevant questions} and many demonstrations fail to meet this standard.

The NISQ hardware available today limits demonstrations to small basis set calculations on small molecules that do not challenge the classical state-of-the-art. Continued hardware development should enable a sequence of demonstrations of increasing size whose end point is quantum advantage. Quantum simulations that address small-scale versions of problems whose large-scale realization is believed to be challenging or out-of-reach of classical chemistry methods are therefore good targets. One may evaluate progress in NISQ demonstrations by considering various desiderata.

It is prudent to question whether a particular NISQ calculation is strictly quantum mechanical in some fundamental conceptual sense. VQE instances have been evaluated from the point of view of contextuality~\cite{kirby2019contextuality}, relevant to the present work. We employ a hybrid quantum-classical method, in which part of the result is solved classically \cite{kirby2020classical} and a quantum correction is calculated on a NISQ device \cite{kirby2021contextual}. This reduces the quantum resource requirements, enabling us to address larger problems and also ensures the quantum calculation is not susceptible to description by some conceptually classical model.

For small problem instances, it is of interest to evaluate the performance of NISQ devices against various classical heuristics.
A standard benchmark problem for many conventional quantum chemistry techniques is molecular nitrogen~\ce{N2}~\cite{lee1993dissociation,van2000benchmark}, which is of particular interest during bond-breaking. \hl{Density matrix renormalization group (DMRG) and coupled cluster calculations were performed on~\ce{N2} in the Dunning cc-pVDZ basis set \cite{chan2004state}, and more recently using heat-bath (HCI) and quantum-selected (QSCI) configuration interaction \cite{robledo2024chemistry}}. In the dissociation limit static correlation dominates \cite{van2000benchmark} and single-reference methods such as Restricted Open-Shell Hartree-Fock (ROHF) break down; in this regime, the ground state wavefunction is not well-described by a single Slater determinant. Despite the inadequacy of the single-reference state, in the limit of all excitations post-Hartree-Fock methods such as Møller–Plesset Perturbation Theory (MP), Configuration Interaction (CI) and Coupled Cluster (CC) are still exact; however, each method requires truncation to be computationally feasible, which induces error. Furthermore, perturbation and coupled cluster approximations suffer from non-variationality \cite[p.~292,~320]{szabo2012modern}, which is observed in the minimal STO-3G basis for the \ce{N2} potential energy curve (PEC) in Figure \ref{fig:final_PEC}.

In such scenarios, multiconfigurational methods are commonly utilized such as complete-active-space configuration interaction (CASCI) and self-consistent field (CASSCF) \cite{roos1980complete}, which account for all determinants that correlate electrons in a specified number of active orbitals and thus have the flexibility to describe mixing between nearly degenerate configurations (i.e. static correlation)~\cite{levine2021cas}. \hl{In Figure \ref{fig:final_PEC} we include CASCI/CASSCF calculations, in each case selecting the active space from MP2 natural orbitals; an occupation number close to zero or two indicates the corresponding spatial orbital is mostly unoccupied/occupied and can therefore be considered inactive, which naturally maximizes the correlation entropy of the wavefunction in the active space. This yields improved treatment of the bond-breaking behaviour for active spaces $(6\mathrm{o},6\mathrm{e})$ and $(7\mathrm{o},8\mathrm{e})$, while coupled cluster is more accurate around the equilibrium geometry where is is expected to perform favourably.} An issue with these CAS methods is that the computational cost scales exponentially with the size of the active space and dynamical correlations outside of the active space are excluded. The missing dynamical correlation can be recovered, for example through low-order perturbations such as complete-active-space (CASPT2) or $n$-electron valence state (NEVPT2) second-order perturbation theory. A further problem with all these techniques is that the quality of the calculation, namely the amount of correlation energy captured, is substantially affected by the choice of active space \cite{stein2016automated}, while keeping the problem computationally tractable.

Another commonly used approach to treating bond-breaking is Unrestricted Hartree-Fock (UHF), in which spin-up and spin-down orbitals are addressed separately. Sometimes, this can qualitatively describe bond dissociation; however, solutions no longer exhibit the correct spatial/spin symmetry \cite{krylov1998size}, i.e. they are no longer eigenstates of the $S^{2} = ||\bm{S}||^2$ operator where $\bm{S} = (S_x, S_y, S_z)$ describes the axial spin components. Since the molecular wavefunction is important to obtain observables other than energy, this represents a drawback of UHF as spin-contaminated or symmetry-broken wavefunctions are inappropriate in such cases.

In this work we invoke the Contextual Subspace approach \cite{kirby2021contextual, weaving2023stabilizer, ralli2023unitary} to quantum chemistry running on superconducting devices. \hl{While we employed this technique previously for the equilibrium ground state preparation of \ce{HCl} on noisy hardware \cite{weaving2023benchmarking}, the variational circuit was preoptimized classically. One other work utilized the Contextual Subspace method on noisy hardware for the purposes of testing a pulse-based ansatz by calculating equilibrium energies \cite{liang2023spacepulse}. However, only the smallest of their simulations, \ce{NH} STO-3G in a $4$-qubit subspace, was able to recover the Hartree-Fock energy, with correlated wavefunction methods a more challenging target. By contrast, in this work we aim to calculate the entire PEC of \ce{N2} -- not just a single point estimate -- with the Contextual Subspace Variational Quantum Eigensolver (CS-VQE) running on a quantum computer; each VQE routine consists of many state preparation and gradient calculations.}

\hl{We compare our methodology against the following conventional quantum chemistry techniques: ROHF, MP2, CISD, CCSD, CCSD(T), CASCI and CASSCF.} Given that we describe the \ce{N2} system in a minimal STO-3G basis set, chosen for its modest size and feasibility on the available quantum hardware, the exact Full Configuration Interaction (FCI) energy can be calculated. The goal here is not to achieve quantum advantage, which is precluded by the limited basis set, but rather to demonstrate how the Contextual Subspace approach on superconducting hardware can be competitive with and challenge a set of classical benchmarks. This framework is fully scalable and, as quantum computers mature, should provide a realistic path to quantum advantage and facilitate practical quantum simulations at large scales.


To realize this goal, we designed a robust simulation methodology that combines various quantum resource reduction tools together with a flexible error mitigation/suppression strategy. An overview of the qubit subspace framework is given in Section \ref{sec:subspace}, while measurement reduction is achieved through a qubit-wise commuting (QWC) decomposition of the reduced Hamiltonians, provided explicitly in Appendix \ref{sec:hamiltonians}. Furthermore, to demonstrate the compatibility of Contextual Subspace with contemporary simulation techniques, we construct ans\"atze via our modification to qubit-ADAPT-VQE \cite{tang2021qubit} which enforces hardware-awareness through a penalising contribution in the excitation pool scoring function, presented in Section \ref{sec:ansatz_construction}. \hl{Details pertaining to the quantum error mitigation/suppression techniques employed for this simulation are provided in Section \ref{sec:QEM}, while software/hardware considerations are outlined in Section \ref{sec:methods}. The paper culminates in Section \ref{sec:results} with quantum computational results for the \ce{N2} PEC.}

\section{Contextual Subspace}\label{sec:subspace}
Contextuality gives us perhaps the broadest conceptual picture of quantum correlations that defy classical description \cite{mermin1990simple, mermin1993hidden, spekkens2007evidence, spekkens2008negativity}. In the restricted setting of Pauli measurements, it manifests in the non-transitivity of commutation amongst non-symmetry elements of the Hamiltonian, implying outcomes may not be assigned consistently without contradiction \cite{kirby2019contextuality, raussendorf2020phase}. Conversely, the absence of contextuality gives way to a class of Hamiltonians whose spectra are described by a classical objective function that parametrizes an underlying hidden-variable model \cite{kirby2020classical}. Therefore, partitioning the target Hamiltonian into contextual and noncontextual components gives us a hybrid quantum/classical algorithm for calculating eigenvalues with reduced quantum resources. In fact, by enforcing noncontextual symmetries over the contextual Hamiltonian, we may identify a so-called \textit{contextual subspace} \cite{kirby2021contextual}.

The qubit reduction mechanism in the contextual subspace approach \cite{kirby2021contextual, weaving2023stabilizer, ralli2023unitary}, also in Qubit Tapering~\cite{bravyi2017tapering, setia2020reducing}, is effected by the stabilizer subspace projection framework \cite{weaving2023stabilizer}; scalable implementations of these techniques are available through the \textit{Symmer} Python package \cite{symmer2022}. Such subspace methods exploit various symmetries of the problem Hamiltonian -- \textit{physical} in the case of Qubit Tapering and \textit{artificial} for Contextual Subspace -- to yield a reduced effective Hamiltonian, with correspondingly reduced quantum resource requirements as measured by number of qubits, number of Hamiltonian terms and Hamiltonian norm, the latter of which dictates the sampling overhead in VQE.

Since Qubit Tapering operates on $\mathbb{Z}_2$-symmetries of the full system, the full and projected Hamiltonians are isospectral up to a change of eigenvalue multiplicities. Indeed, degeneracies of the full system may be lifted under this procedure; any remaining degeneracy implies the existence of non-$\mathbb{Z}_2$ symmetry. All molecular systems possess at least two $\mathbb{Z}_2$-symmetries which enforce the parity of spin up or down particles. Additional symmetry arises from the molecular point group that describes the geometrical symmetry of the system. In the setting of diatomic molecules there are two relevant point groups: the cyclic point group $C_{\infty v}$ consisting of continuous rotations around the inter-nuclear axis and the dihedral point group $D_{\infty h}$ that also includes the reflection and inversion symmetries of the diatomic. Heteronuclear molecules, consisting of two distinct atomic centres, lie within the former group, while homonuclear molecules such as \ce{N2} fall under the latter.

The specific $\mathbb{Z}_2$-symmetries one exploits through tapering come from abelian subgroups of the above point groups that describe a restriction to 2-fold symmetry. In particular, the relevant group generators of $C_{2v} \subset C_{\infty v}$ are $180^{\circ}$ rotations around the bond axis, denoted $C_2$, and vertical reflections $\sigma_{v}$. In the case of $D_{2h} \subset D_{\infty h}$ we have the same rotational symmetry $C_2$, in addition to the group generators $\sigma_h$, corresponding with horizontal reflections, and the inversion symmetry $i$. In all, qubit tapering enables the removal of four qubits from heteronuclear molecules (two point group generators $C_2, \sigma_v$ plus spin up/down parity) and five from homonuclear molecules (three point group generators $C_2, \sigma_h, i$ plus spin up/down parity). \hl{This reduces our \ce{N2} Hamiltonian from 20 to 15 qubits without incurring any error.}

The Contextual Subspace methodology is more general than Qubit Tapering. The reduced Hamiltonians produced via this technique need not preserve the spectrum of the full Hamiltonian, which makes the selection of stabilizers (one might think of these as `pseudo'-symmetries) that define the subsequent contextual subspace more nuanced. \hl{We motivate this selection from the MP2 wavefunction (specifically, the excitation generator) just as in the CASCI and CASSCF calculations so that the methods may be compared fairly. While the latter selects an active space informed by the MP2 natural orbitals, we maximize the $\ell_1$-norm of the MP2 expansion coefficients whose corresponding Pauli-terms commute with the chosen stabilizers.}

\begin{figure}[t]
    \centering
    \includegraphics[width=\linewidth]{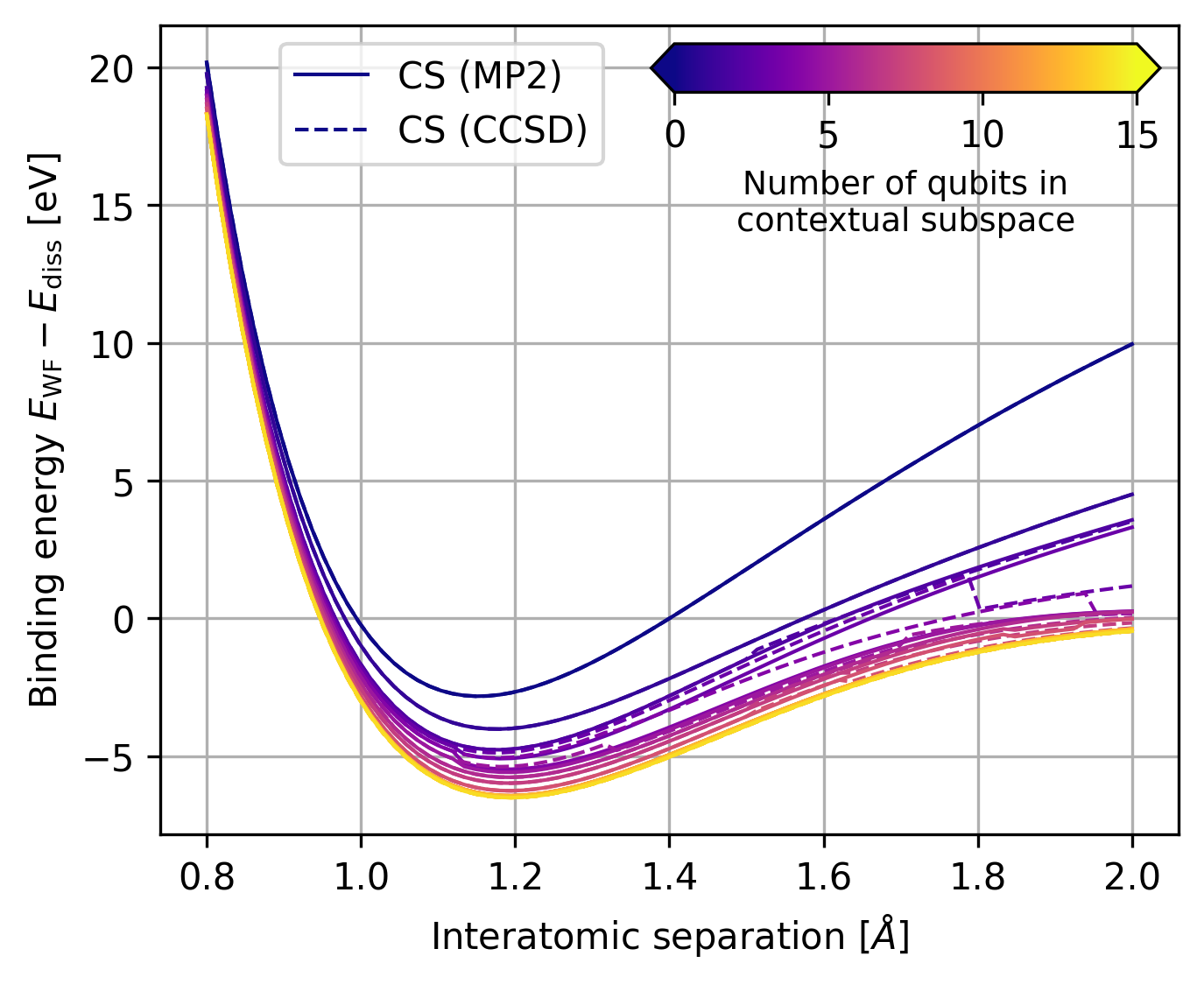}
    \caption{\hl{Binding potential energy curves in contextual subspaces whose corresponding stabilizers are informed either by the MP2 or CCSD wavefuntions. In the latter we encounter a large number of discontinuities that are largely absent for the MP2-informed subspaces, although the absolute error is typically lower. At 15-qubits we recover the full space and the exact FCI energy is obtained.}}
    \label{fig:CS_MP2_CCSD_comp}
\end{figure}

\hl{In the development of this work we also compared subspaces motivated by the CCSD excitation generators; interestingly, we found discontinuities in the resulting PEC, which may be observed in Figure \ref{fig:CS_MP2_CCSD_comp}. This was encountered not just in the contextual subspace, but also for the CASCI/CASSCF calculations as seen in Figure \ref{fig:CASCI_disc} of Appendix \ref{sec:CASCI_active_space}. Hence, we deferred to the MP2 wavefunction where such discontinuities were largely alleviated, at the cost of the absolute error generally being greater than that found in the CCSD-motivated subspaces.}

\begin{figure}[b]
    \centering
    \includegraphics[width=\linewidth]{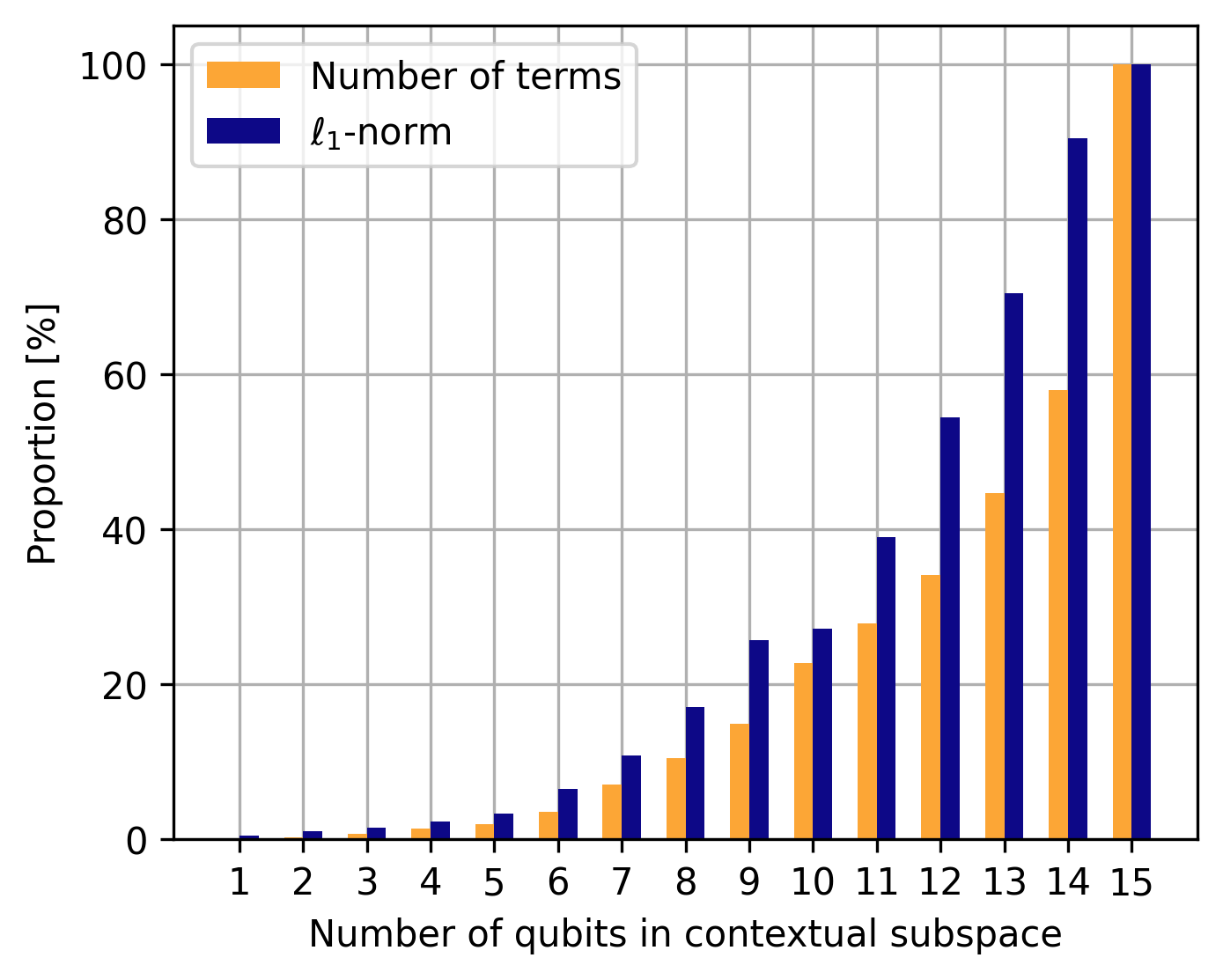}
    \caption{\hl{Proportion of the number of terms and $\ell_1$-norm of the reduced contextual subspace Hamiltonians versus the full system at the equilibrium bond length $r=1.192\angstrom$.}}
    \label{fig:CS_proportions}
\end{figure}

We may construct contextual subspace approximations for any number of qubits between $1 - 14$, given that we first taper the molecule so a contextual subspace on $15$ qubits corresponds with performing full-system VQE. \hl{Not only does the contextual subspace method allow us to reduce the number of qubits required to represent a Pauli Hamiltonian, it also has an impact on the number of terms and $\ell_1$-norm of the resulting Pauli coefficients, $\Lambda = \sum_i |h_i|$, as seen in Figure \ref{fig:CS_proportions}. This has implications on the sampling overhead required in VQE, which scales asymptotically as $\mathcal{O}(\Lambda^2\epsilon^{-2})$ to achieve a desired error $\epsilon>0$~\cite{wecker2015progress, rubin2018application}.}

\hl{It can be argued that reducing the number of qubits not only allows us to simulate larger systems on quantum hardware, but may also render such systems classically-tractable. However, the fact that $\Lambda$ is additionally reduced, thus alleviating the quantum overhead further, provides a strong motivation for its use as a quantum resource reduction technique. In addition, this feature will benefit Hamiltonian simulation techniques such as QDRIFT~\cite{campbell2019random}, where there is an explicit quadratic scaling dependence on $\Lambda$ in the resulting circuit depths.}

\hl{The contextual subspace approximation is also compatible with more advanced measurement-reduction methodologies; in previous work we studied its use in combination with unitary partitioning~\cite{ralli2023unitary}, although we did not implement it for this experiment as a unitary must be applied in-circuit to realize the measurement of each anticommuting clique \cite{izmaylov2019unitary, zhao2020measurement, ralli2021implementation}. If combined with additional techniques for reducing the number of terms in the Hamiltonian, such as tensor hypercontraction~\cite{lee2021even}, this could present a compelling quantum resource management framework.}

Increasing the number of qubits in the contextual subspace increases the accuracy of the method. For \ce{N2}, \hl{in order to achieve algorithmic accuracy (terminology introduced in \cite{motta2022emerging} and taken here to mean errors within $43$~meV of FCI, with \textit{chemical} accuracy a common misnomer when working within minimal basis sets since it implies agreement with experimental results) throughout the full PEC under the contextual subspace approximation, we need 11/12-qubits motivated by the CCSD/MP2 wavefunctions, respectively. However, increasing the number of qubits in the contextual subspace also increases the depth of the ansatz circuit and hence exposes us to the vulnerabilities of hardware noise. In Figure \ref{fig:linearity_biasing} of the following section we present the results of running noiseless qubit-ADAPT-VQE \cite{tang2021qubit, grimsley2019adaptive} over a 12-qubit subspace and observe the decay of error against the number of CNOT gates in the ansatz circuit; such circuits are too deep to obtain satisfactory results on the available hardware.} There is a trade-off between a sufficiently large contextual subspace to represent the problem accurately and a sufficiently shallow ansatz circuit such that the output is not overly contaminated by noise. 

We have been able to achieve \hl{algorithmic accuracy} on quantum hardware in previous work for the equilibrium ground state of \ce{HCl}~\cite{weaving2023benchmarking}, where just 3-qubits were sufficient and hence a shallow ansatz was possible. A 12-qubit ansatz circuit would be too deep -- and consequently too noisy -- to achieve this level of accuracy on current hardware. Since algorithmic accuracy is too challenging a target for a 12-qubit simulation on real hardware, we relax this requirement. \hl{Instead we choose a contextual subspace that is sufficiently large to challenge a set of classical methods for \ce{N2}. We compare against ROHF, MP2, CISD, CCSD, CCSD(T), CASCI and CASSCF for active spaces of varying size.} In reproducing the PEC of \ce{N2} we find that a 5-qubit contextual subspace, while not \hl{algorithmically accurate, yields errors that do not exceed $1$~eV, as shown in Figure \ref{fig:final_PEC}.} It should be highlighted that the above classical techniques do not maintain algorithmic accuracy throughout the PEC either.
\section{Ansatz Construction}\label{sec:ansatz_construction}
We adopt qubit-ADAPT-VQE \cite{tang2021qubit, grimsley2019adaptive} to build sufficiently shallow circuits for our \ce{N2} PEC experiment, with a new modification that facilitates a hardware-aware approach to adaptive circuit construction. The general ADAPT framework is provided in Algorithm \ref{adapt_alg}; the central component is a pool of Pauli operators $\mathcal{P}$, from which one builds an ansatz circuit $\ket{\psi}$ iteratively by appending the term that maximizes some scoring function $f$ at each step. In the standard approach we take the partial derivative at zero after appending a given pool element $P \in \mathcal{P}$, specifically
\begin{equation}\label{standard_ADAPT_score}
    f(P) \coloneqq \frac{\partial}{\partial \theta} \bra{\psi} e^{- i \theta P} H e^{i \theta P} \ket{\psi} \big|_{\theta = 0},
\end{equation}
which may be evaluated either with the parameter shift rule \cite{parrish2019hybrid} or by measuring the commutator $[H,P]$ \cite{grimsley2019adaptive}. By calculating the pool scores $f(\mathcal{P})$ and identifying the maximal term, we extend $\ket{\psi} \rightarrow e^{i \theta P} \ket{\psi} $ and re-optimize the ansatz parameters via regular VQE before repeating. For our particular application, we take $\mathcal{P}$ to be the set of single and double coupled-cluster excitations.

We modify the pool scoring function \eqref{standard_ADAPT_score} to enforce hardware-awareness in the adaptive circuit construction, \hl{thus minimizing the number of SWAP operations incurred through transpilation. We achieve this by ranking approximate subgraph isomorphisms in the hardware topology, described by a graph $\mathcal{G}_{\mathrm{target}} = (\mathcal{N}_{\mathrm{target}}, \mathcal{E}_{\mathrm{target}})$ where $\mathcal{N}_{\mathrm{target}}$ is the set of available qubits and $\mathcal{E}_{\mathrm{target}} \subset \mathcal{N}_{\mathrm{target}}^{\times 2}$ is the edge-set indicating that two qubits may be natively coupled via some nonlocal operation on the hardware. We define an isomorphism between two graphs $\mathcal{G}$ and $\mathcal{H}$ to be a bijective map $g:\mathcal{N}_{\mathcal{G}} \mapsto \mathcal{N}_{\mathcal{H}}$ such that, if $(u,v) \in \mathcal{E}_{\mathcal{G}}$, then $(g(u),g(v)) \in \mathcal{E}_{\mathcal{H}}$. In other words, an isomorphism is a mapping from nodes of $\mathcal{G}$ onto nodes of $\mathcal{H}$ that preserves the adjacency structure. Furthermore, two graphs are said to be \textit{subgraph} isomorphic if $\mathcal{G}$ is isomorphic to a subgraph of $\mathcal{H}$; we use the $\mathrm{VF2^{++}}$ algorithm \cite{juttner2018vf2++} as implemented in the \textit{NetworkX} Python package \cite{hagberg2008exploring} for subgraph isomorphism matching.}

\hl{In order to reweight the standard score assigned to a given pool operator $P \in \mathcal{P}$, we construct a weighted graph $\mathcal{G}_{\mathrm{circuit}} = (\mathcal{N}_{\mathrm{circuit}}, \mathcal{E}_{\mathrm{circuit}})$ for the circuit $e^{i\theta P} \ket{\psi}$ and bias with respect to a notion of distance from the nearest subgraph isomorphism, described in Algorithm \ref{topology_aware_bias}. This works by iteratively deleting collections of qubits $\bm{n} \in \mathcal{N}_{\mathrm{circuit}}^{\times d}$ from the ansatz circuit and any associated edges in the corresponding coupling graph, terminating once a subgraph isomorphism is identified. Here, $d$ is the search-depth, which begins at $d=0$ with no qubits deleted and is incremented at each step; since the number of distinct $\bm{n}$ is ${|\mathcal{N}_{\mathrm{circuit}}| \choose d}$, we truncate at some maximum depth $D$ and any pool operator for which no subgraph isomorphism was found with $d \leq D$ receives a score of zero. Otherwise, with the function $s(\bm{n})$ that sums edge-weights connected to the nodes $\bm{n}$, our new Hardware-Aware ADAPT-VQE scoring function becomes
\begin{equation}\label{new_score}
    f(P) \coloneqq \bigg(1-\frac{s(\bm{n})}{W}\bigg)^b \cdot \frac{\partial}{\partial \theta} \bra{\psi} e^{- i \theta P} H e^{i \theta P} \ket{\psi} \big|_{\theta = 0}
\end{equation}
where $W=\sum_{(u,v,w) \in \mathcal{E}_{\mathrm{circuit}}}w$ is the total sum of edge-weights and $b>0$ is the biasing strength. This allows one to control the severity with which non-subgraph-isomorphic circuits are penalised. While the depth $d$ does not explicitly appear in equation \eqref{new_score}, since $|\bm{n}| = d$ we will have more edge-weights included in $s(\bm{n})$ for larger depths and hence will be penalized more.}

\begin{figure}[b]
    \includegraphics[width=\linewidth]{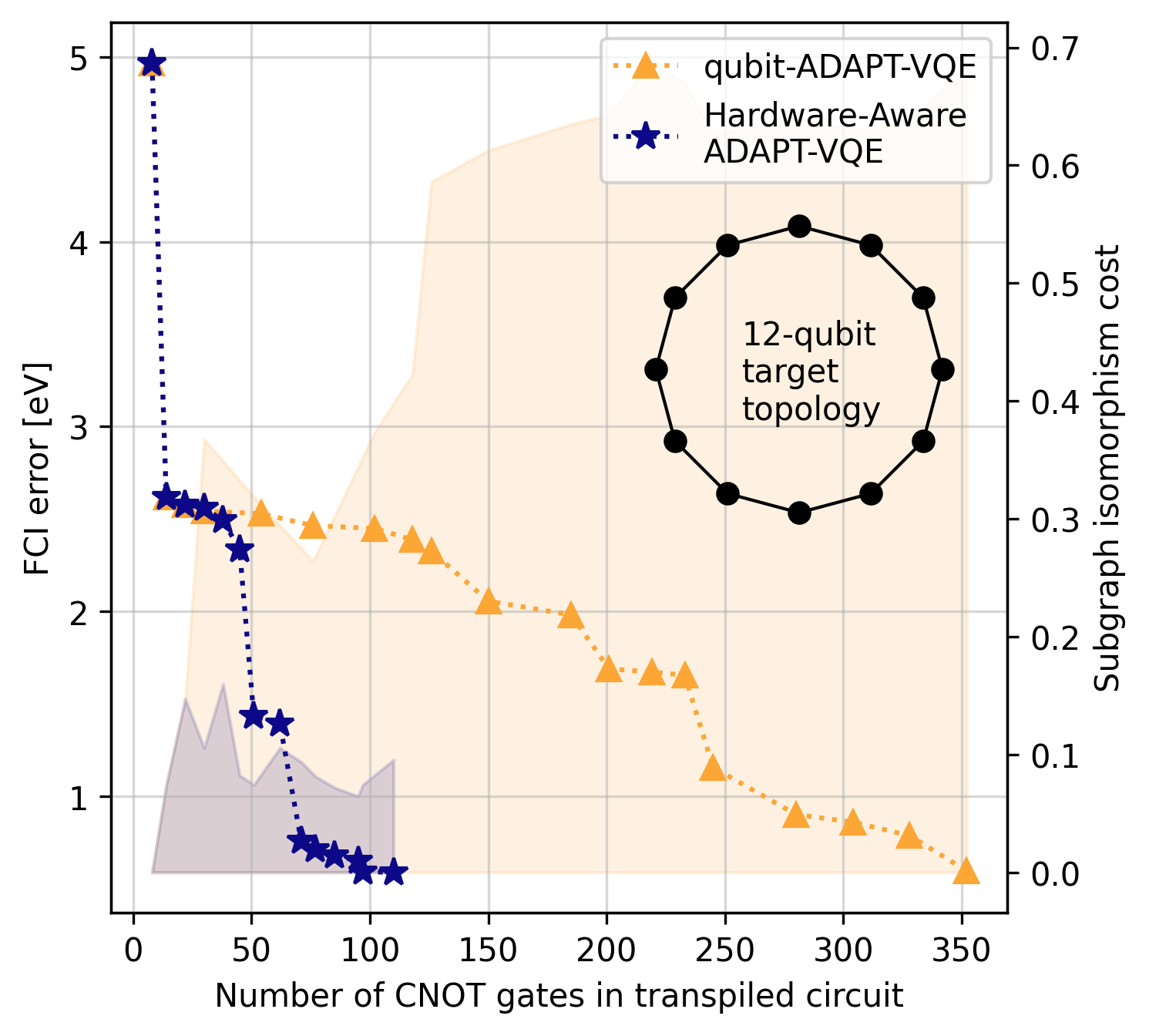}
    \caption{\hl{Construction of a 12-qubit contextual subspace ansatz for \ce{N2} at $2\angstrom$. We show the FCI error per ADAPT cycle against the number of CNOT operations in the corresponding circuits transpiled for a closed loop of the 27-qubit \textit{Falcon} topology as in Figure \ref{fig:circuit_tiling}. We compare the standard qubit-ADAPT-VQE algorithm versus our hardware-aware approach and observe considerably reduced depths. The `subgraph isomorphism cost' of embedding the ansatz graph in the target is computed as $s(\bm{n})/W$.}}
    \label{fig:linearity_biasing}
\end{figure}

\hl{We test our new hardware-aware ADAPT objective function by constructing a 12-qubit contextual subspace ansatz for \ce{N2} at a stretched bond length of $2\angstrom$. For the target topology we choose a 12-qubit ring, which is found as a subgraph of the Falcon chip layout presented in Figure \ref{fig:circuit_tiling}. In Figure \ref{fig:linearity_biasing} we compare error against the number of CNOT gates in the transpiled circuit for our new scoring function, versus the standard qubit-ADAPT-VQE approach. Transpilation is the mapping of a given circuit onto the target quantum device, which may not natively support the required entangling operations and thus expensive SWAP operations are incurred to compensate for discrepancies in the qubit connectivity. The number of two-qubit gates required to transpile the ansatz circuit for the chosen 12-qubit ring is seen to be dramatically reduced, while maintaining similar errors compared with the hardware-agnostic approach. For fairness, both techniques were transpiled using a basic level of circuit optimization (e.g. cancellation of inverse gates).}

\hl{Details for the ansatz construction algorithms may be found in Appendix \ref{sec:adapt_algos}, while in Appendix \ref{sec:ansatz_circuits} we present the 5-qubit quantum circuits produced via hardware-aware ADAPT-VQE that were subsequently executed on IBM hardware to produce the results of Section \ref{sec:results}. We note that the circuits presented there have received a high level of optimization in order to yield the lowest possible depth and are not the untreated circuits obtained directly from our ansatz construction routine.}
\section{Quantum Error Mitigation/Suppression}\label{sec:QEM}
\hl{The handling of errors in quantum computation can be categorized as suppression, mitigation or correction. The first of these is implemented close to the hardware itself and attempts to deal with flaws in the operation and control of the device. Mitigation, on the other hand, serves to reduce bias in some statistical estimator of interest by executing ensembles of circuits that have been carefully designed to exploit a feature of the quantum noise; this typically comes at the cost of increased uncertainty in the resulting expectation values. Finally, error correction schema engineer redundancy into the system, forming `logical qubits' from many physical qubits such that errors may be detected and corrected on-the-fly during computation.}

\hl{The error handling strategy deployed for this work deploys methods of suppression and mitigation} and is motivated by the results of our previous benchmarking effort, in which we estimated the ground state energy of the \ce{HCl} molecule to algorithmic accuracy (within $43$ meV of FCI) \cite{weaving2023benchmarking}. For our \ce{N2} simulations, we adopt a combination of dynamical decoupling (\ref{sec:DD}) and measurement-error mitigation (\ref{sec:MEM}) with zero-noise extrapolation (\ref{sec:ZNE}). In our previous work we were also able to exploit Symmetry Verification \cite{bonet2018low, mcardle2019error, cai2021quantum}, however the non-$\mathbbm{Z}_2$-symmetries (particle and spin quantum number) reduce to the identity under our qubit subspace procedure in this case and therefore do not permit error mitigation opportunities. We also employ a circuit parallelization scheme that averages over hardware noise (\ref{PNA}).

\subsection{Dynamical Decoupling}\label{sec:DD}

The original mechanism underpinning the Dynamical Decoupling (DD) \hl{error suppression technique}, in which a carefully applied sequence of pulses may prolong the coherence of a spin system, predates quantum computing \cite{hahn1950spin}. In our case, we apply periodic spin echos on idling qubits to suppress undesirable coupling between the system and its environment~\cite{viola1998dynamical, facchi2005control}. We use a simple uniform sequence of $\pi$-pulses to effect the decoupling; different sequences with non-uniform spacing (such as Uhrig DD \cite{uhrig2007keeping}) might yield improvements.

\subsection{Measurement-Error Mitigation}\label{sec:MEM}

In Measurement Error Mitigation (MEM) we apply an inverted transition matrix representing the probability of a bitflip $\ket{0} \leftrightarrow \ket{1}$ occurring for a given qubit \cite{bravyi2021mitigating}. More specifically, for a qubit $k$ we evaluate a $2 \times 2$ matrix where element $A_{ij}^{(k)}$ represents the probability of preparing state $\ket{i} \in \{\ket{0},\ket{1}\}$ and measuring state $\ket{j} \in \{\ket{0},\ket{1}\}$. Tensoring qubits together allows us to infer the joint probabilities of preparing $\ket{\bm{i}} \in \{\ket{0},\ket{1}\}^{\otimes N}$ and measuring $\ket{\bm{j}} \in \{\ket{0},\ket{1}\}^{\otimes N}$ by taking products over the individual qubit marginals 
\begin{equation}
A_{ij} \approx \prod_{k=0}^{N-1} A^{(k)}_{\bm{i}_k, \bm{j}_k};    
\end{equation}
applying $A^{-1}$ to any subsequent noisy measurement results provides a rectified readout distribution.

This makes a strong assumption that quantum measurements are independent; Nation \textit{et al.} demonstrate a breakdown of this technique in the presence of correlated measurements \cite{nation2021scalable}, however constructing the fully coupled transition matrix is not in general feasible. The \textit{mthree} package \cite{nation2021scalable} was utilized to implement MEM for our simulations. \hl{We note there are alternative techniques that do make such assumptions on the nature of the readout error, such as Twirled Readout Error Extinction (TREX), which leverages the idea of twirled measurements \cite{van2022model}.}

\subsection{Circuit Tiling}\label{PNA}

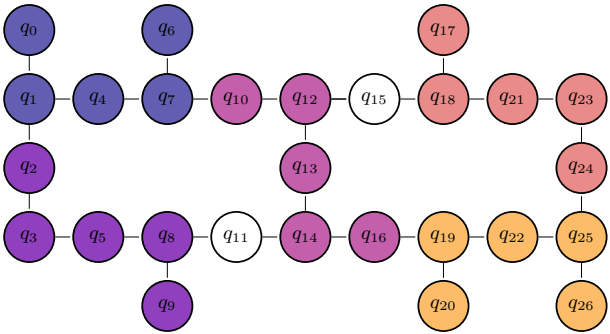
\begin{figure}[h]

        \resizebox{0.95\linewidth}{!}{
        \begin{tikzpicture}[shorten >=1pt, auto, node distance=11mm,
         every node/.style={circle,thick,draw,minimum size=8mm}
         ]
        \node[fill=plasma0!65] (A)              {$q_0$};    \node[fill=plasma0!65] (B) [ below of=A] {$q_1$};    \node[fill=plasma0!65] (C) [right of=B] {$q_4$};
        \node[fill=plasma0!65] (D) [right of=C] {$q_7$};    \node[fill=plasma2!75] (E) [right of=D] {$q_{10}$}; \node[fill=plasma2!75] (F) [right of=E] {$q_{12}$};
        \node (G) [right of=F] {$q_{15}$}; \node[fill=plasma3!75] (H) [right of=G] {$q_{18}$}; \node[fill=plasma3!75] (I) [right of=H] {$q_{21}$};
        \node[fill=plasma3!75] (J) [right of=I] {$q_{23}$};
        \node[fill=plasma1!75] (K) [ below of=B] {$q_2$};
        \node[fill=plasma1!75] (L) [ below of=K] {$q_3$};    \node[fill=plasma1!75] (M) [right of=L] {$q_5$};    \node[fill=plasma1!75] (N) [right of=M] {$q_8$};
        \node (O) [right of=N] {$q_{11}$}; \node[fill=plasma2!75] (P) [right of=O] {$q_{14}$}; \node[fill=plasma2!75] (Q) [right of=P] {$q_{16}$};
        \node[fill=plasma4!75] (R) [right of=Q] {$q_{19}$}; \node[fill=plasma4!75] (S) [right of=R] {$q_{22}$}; \node[fill=plasma4!75] (T) [right of=S] {$q_{25}$};
        \node[fill=plasma4!75] (U) [ below of=T] {$q_{26}$};
        \node[fill=plasma3!75] (V) [above of=T] {$q_{24}$}; \node[fill=plasma2!75] (W) [ below of=F] {$q_{13}$}; \node[fill=plasma1!75] (X) [ below of=N] {$q_9$};
        \node[fill=plasma4!75] (Y) [ below of=R] {$q_{20}$}; \node[fill=plasma0!65] (Z) [above of=D] {$q_6$}; \node[fill=plasma3!75] (ZA)[above of=H] {$q_{17}$};
        \draw (A) edge (B); \draw (B) edge (C); \draw (C) edge (D); \draw (D) edge (E); \draw (E) edge (F); 
        \draw (F) edge (G); \draw (G) edge (H); \draw (H) edge (I); \draw (I) edge (J); \draw (F) edge (G);
        \draw (B) edge (K); \draw (K) edge (L); \draw (L) edge (M); \draw (M) edge (N); \draw (N) edge (O);
        \draw (O) edge (P); \draw (P) edge (Q); \draw (Q) edge (R); \draw (R) edge (S); \draw (S) edge (T);
        \draw (T) edge (U); \draw (T) edge (V); \draw (V) edge (J); \draw (F) edge (W); \draw (D) edge (Z);
        \draw (P) edge (W); \draw (N) edge (X); \draw (R) edge (Y); \draw (H) edge (ZA);
        \end{tikzpicture}
        } 
    \label{clusters}
    
    \caption{An example of circuit tiling over the IBM 27-qubit `heavy-hex' topology found in their \textit{Falcon} series chips. The different colours indicate replica ansatz circuits tiled across 5-qubit clusters. Not only does this increase the effective number of measurements extracted from the hardware 5-fold, but it also serves as a form of passive quantum error mitigation whereby noise is averaged over the device.}
    \label{fig:circuit_tiling}
\end{figure}

Noise is not uniform across the qubits of a quantum processor, hence one will observe considerable variance in the results when executing the same circuit on different parts of the chip; to mitigate this, we execute many replica circuits across the chip and average over the results, which has the added benefit of increasing the effective shot-yield. We depict our circuit parallelization scheme in Figure \ref{fig:circuit_tiling} for the 27-qubit \textit{Falcon} architecture, which extends similarly to the 127-qubit \textit{Eagle} device. One may view this as instance of ensemble averaging, often employed when computational models exhibit severe sensitivity to the initial conditions.

This noise averaging process results in improved stability of the final energy estimates, especially when used in combination with DD and MEM as introduced above. This is particularly desirable if one wishes to make inferences from the behaviour of these estimates under some noise amplification procedure, which is precisely the case for ZNE, introduced in the following Section \ref{sec:ZNE}. When performing regression, any uncertainty in the extrapolation data will propagate through to the inferred values and thus increase the variance of the final energy estimate. This has also been observed when applying the error mitigation technique of randomized compilation (RC) \cite{wallman2016noise} in combination with ZNE, where it is argued that small amounts of coherent error lead to substantial errors \cite{kurita2022synergetic}. While RC converts coherent error into stochastic Pauli noise by implementing the target unitary in different ways, one might draw an analogy with the parallelization scheme presented here. Indeed, due to inconsistency in the noise channels for local qubit clusters, the unitary performed in each sub-circuit will differ and might explain the stable noise amplification observed in Figures \ref{fig:noise_fitting} and \ref{fig:noise_extrapolation}, ultimately leading to reliable extrapolation and lower variance in the final energy estimate.

\subsection{Zero-Noise Extrapolation}\label{sec:ZNE}

Zero-Noise Extrapolation (ZNE), also referred to as Richardson Extrapolation, is a heuristic that amplifies noise arising from some selection of gates in the quantum circuit, obtaining noise-scaled energy estimates that one may use to extrapolate to the hypothetical point of `zero-noise' \cite{li2017efficient, temme2017error, endo2018practical, kandala2019error, giurgica2020digital, he2020zero, mari2021extending, maupin2023error, kim2023evidence}. ZNE is sensitive to the way in which one chooses to perform the noise amplification; \hl{some works choose to amplify noise in the temporal domain through gate stretching, requiring pulse-level control of the hardware \cite{temme2017error, kandala2019error}, while others adopt discrete approaches such as unitary folding \cite{giurgica2020digital} where identities $I=U^\dag U$ are injected into the circuit. In all cases} care must be taken to ensure circuit transpilation does not perform any internal circuit optimizations that would lead to unpredictable noise scaling.

\begin{figure}[h]
    \resizebox{0.9\linewidth}{!}{
    \begin{quantikz}[row sep=0.5cm, column sep=0.5cm]
    \lstick{$c$} & \ctrl{1} & \qw \\
    \lstick{$t$} & \targ{} & \qw
    \end{quantikz} 
    $\equiv$
    \begin{quantikz}[row sep=0.4cm, column sep=0.5cm]
    \lstick{$c$} & \qw      & \ctrl{1}\gategroup[2,steps=1,style={dashed,rounded corners,fill=plasma4!50, inner xsep=2pt},background]{{\sc $\lambda$} repetitions}                  &  \qw      & \qw \\
    \lstick{$t$} & \gate{H} & \gate{P(\frac{\pi}{\lambda})} & \gate{H} & \qw
    \end{quantikz}
    }
    \caption{A root-product decomposition of the CNOT gate into CPhase gates, repeated application of which is counteracted with reduced rotation angles.}
    \label{fig:noise_amp_method}
\end{figure}
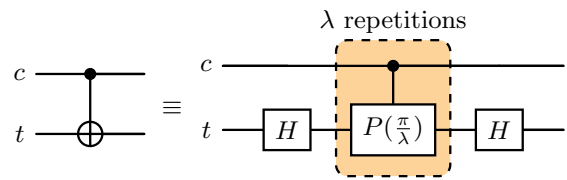

For this experiment, we adopt the strategy used in our previous work \cite{weaving2023benchmarking} to decompose each CNOT operation into a product over its roots, where we may increase the number $\lambda \in \mathbbm{N}$ of factors in the decomposition \hl{
\begin{equation}
\begin{aligned}
\mathrm{CNOT}_{c,t} ={} & \prod_{l=1}^{\lambda} \sqrt[\lambda]{\mathrm{CNOT}_{c,t}}\\ ={} & \mathrm{Had}_t \Big[\prod_{l=1}^{\lambda} \mathrm{CPhase}_{c,t}\Big(\frac{\pi}{\lambda}\Big)\Big] \mathrm{Had}_t
\end{aligned}
\end{equation}
as in Figure \ref{fig:noise_amp_method}. The control and target qubits are denoted $c$ and $t$, respectively, noting also that there are Hadamard gates $\mathrm{Had}_t$ between consecutive $\mathrm{CPhase}$ gates that cancel, leaving only those at the beginning and end.} Given the IBM native gateset, each CPhase gate will need to be transpiled back into a pair of CNOTs and three $R_z$ rotations; thus, each CNOT gate is replaced with $2 \lambda$ CNOT and $3 \lambda + 2$ single-qubit gates through this noise amplification procedure.

\hl{Previously, we experimented with a local unitary folding scheme in which we inserted CNOT pairs after each CNOT in the circuit, resulting in each being replaced with $2\lambda+1$ CNOT gates, compared with the $2\lambda$ encountered in our CPhase approach. In order for $\lambda \rightarrow 0 $ to probe the zero-CNOT-noise regime, we needed to offset the noise amplification/gain factors in the extrapolation to account for the additional $+1$ CNOT of the former. By contrast, we find the CPhase decomposition to avoid the necessity of this gain offset, making for cleaner regression.}

\begin{figure}[t]
    \centering
    \includegraphics[width=\linewidth]{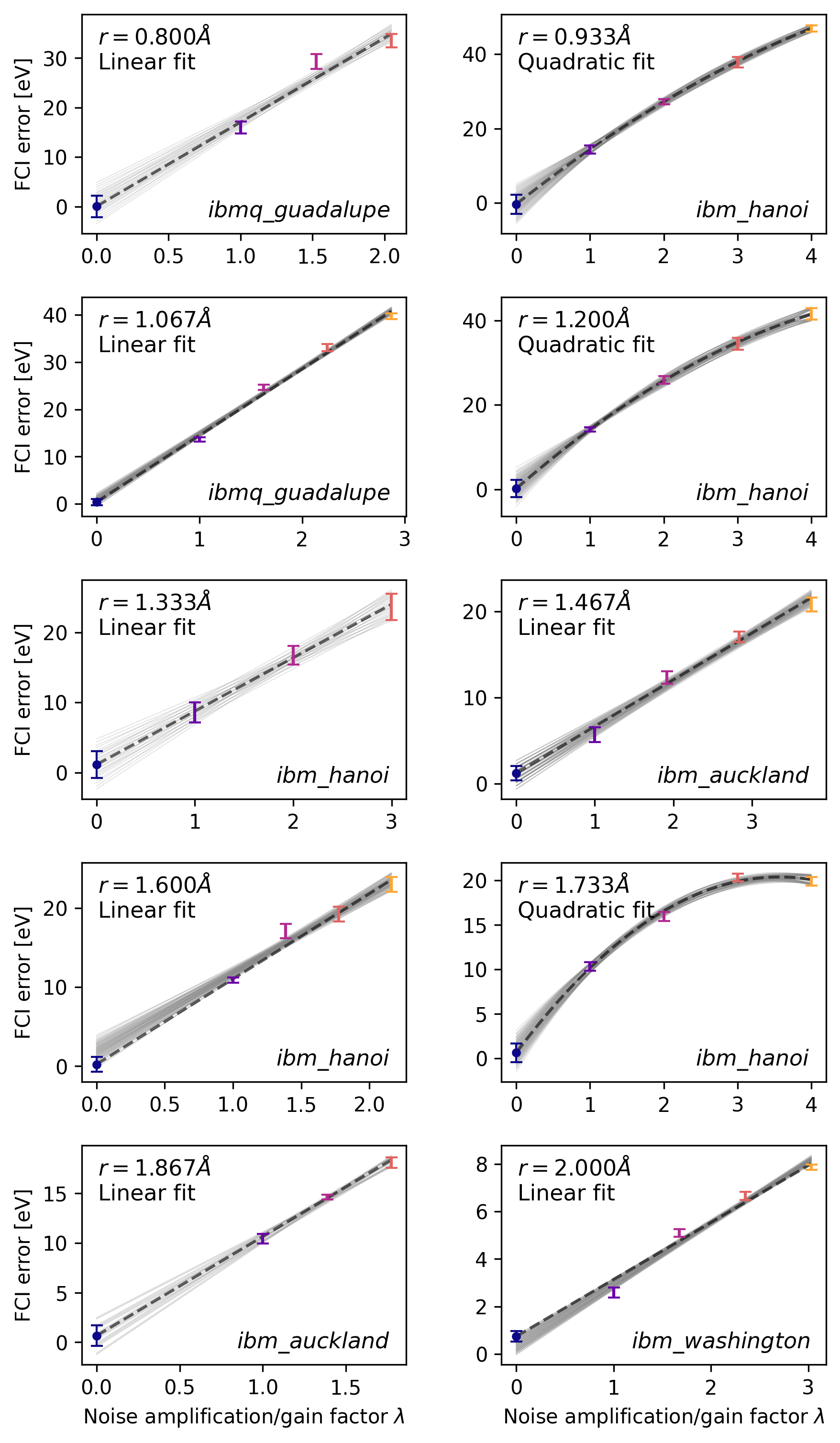}
    \caption{\hl{Noise fitting curves for ten evenly spaced interatomic separations of molecular nitrogen. Standard deviations were taken over the converged data in VQE in order to use weighted least squares in the linear/quadratic regression. We also plotted the spread of possible extrapolation curves given the variance of each individual noise amplified estimate. The noise amplification factors themselves were calibrated using one- and two-qubit gate error data extracted from the hardware at the time of execution.}}
    \label{fig:noise_fitting}
\end{figure}

\begin{figure}[t]
    \centering
    \includegraphics[width=\linewidth]{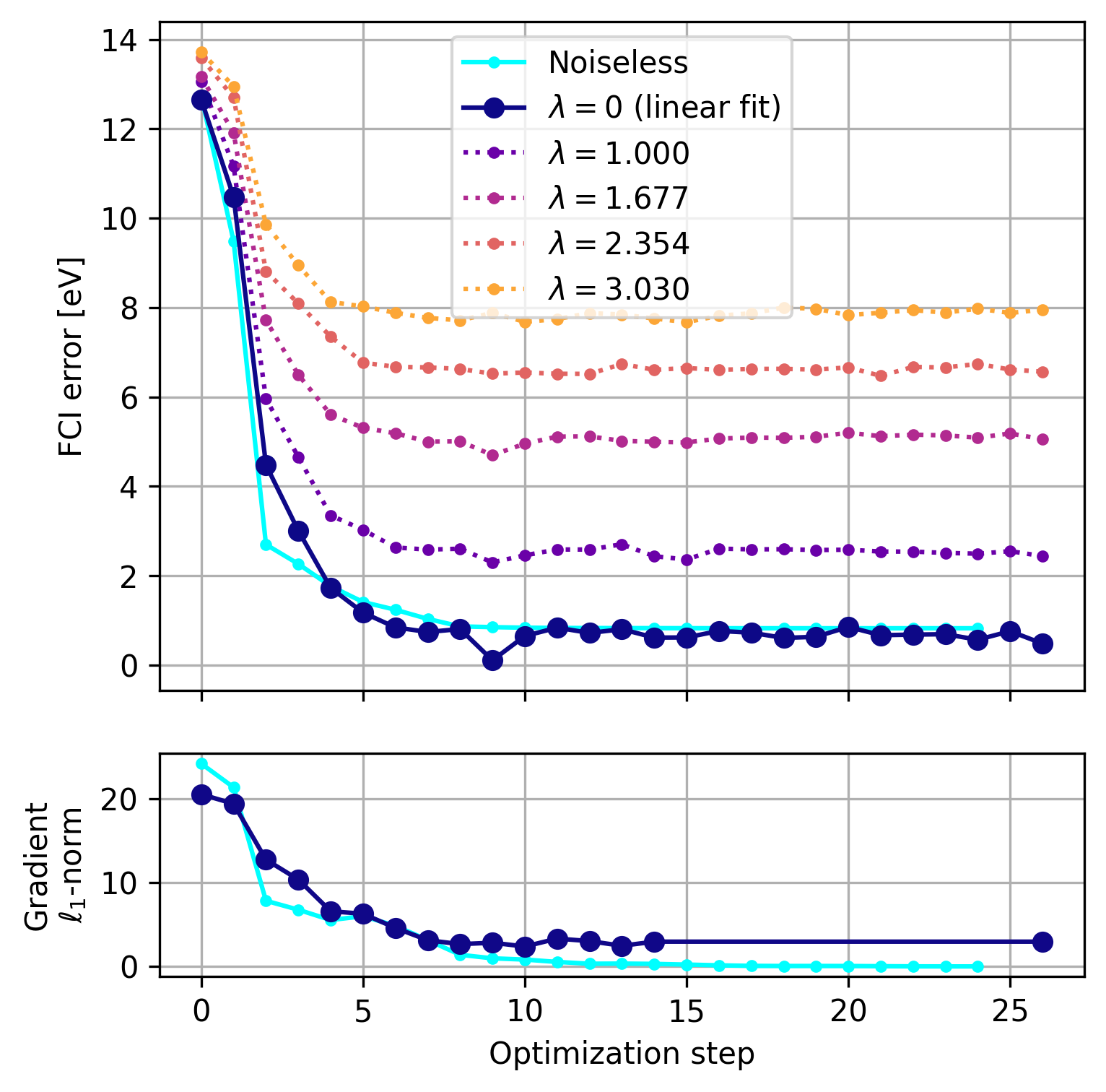}
    \caption{\hl{Noise amplified VQE routine at $r=2\angstrom$ on \textit{ibm\_washington}. We also include a noiseless routine for comparison and note the partial derivatives converge on zero in the noiseless simulation, while they are prevented from doing so in the noisy case indicated by non-zero gradient $\ell_1$-norm, despite the optimizer having converged on the ground state energy.}}
    \label{fig:noise_extrapolation}
\end{figure}

\hl{After retrieving the noise-amplified results from the quantum device, the noise amplification factors are calibrated using the one- and two-qubit gate error data extracted from the hardware at the time of execution; this is why the extrapolation data do not lie on integer values of $\lambda$ in Figure \ref{fig:noise_fitting}. For more reliable extrapolation, we employ inverse variance weighted least squares regression (linear or quadratic), so that highly varying data points are penalised in the fitting procedure; the variances here are obtained from the converged VQE data, as opposed to the statistical bootstrapping procedure we used in our previous ZNE work \cite{weaving2023benchmarking}. In Figure \ref{fig:noise_extrapolation} we present a full VQE routine executed on the \textit{ibm\_washington} system, complete with the noise amplified data that leading to the final extrapolated estimate.}

\hl{\section{Software and Hardware Implementation Details}\label{sec:methods}}
All of the conventional quantum chemistry techniques used within this work were facilitated by \textit{PySCF} \cite{sun2018pyscf}. Hamiltonian construction began with a restricted open-shell Hartree-Fock calculation, before building second-quantized fermionic operators in \textit{OpenFermion} \cite{mcclean2020openfermion} that were subsequently mapped onto qubits via the Jordan-Wigner transformation \cite{jordan1993paulische}. The resulting operators were then converted into \textit{Symmer} objects in order to leverage the included qubit subspace functionality \cite{symmer2022}; we projected each Hamiltonian onto a 5-qubit subspace before running our VQE simulations.

Ansatz circuits were constructed using the Hardware-Aware ADAPT-VQE algorithm of Section \ref{sec:ansatz_construction}, with the optimized circuits given explicitly in Appendix \ref{sec:ansatz_circuits}. We extracted qubit-wise commuting (QWC) cliques from the Hamiltonians using graph-colouring functionality in \textit{NetworkX}~\cite{hagberg2008exploring}, with the explicit decompositions provided in Appendix \ref{sec:hamiltonians}. We performed $5,000$ circuit shots for every QWC clique per expectation value calculation and the classical optimizer in each VQE routine was the Broyden–Fletcher–Goldfarb–Shanno (BFGS) algorithm \cite{fletcher2013practical}. Gradients were calculated in hardware via the parameter shift rule \cite{parrish2019hybrid}. The simulations were executed on 15 of the available 16 qubits on Falcon r4P QPU \textit{ibmq\_guadalupe} (3 ansatz circuit tilings), 25 out of 27 qubits on Falcon r5.11 QPUs \textit{ibm\_hanoi}, \textit{ibm\_auckland} (5 ansatz circuit tilings) and 125 of the 127 qubits on the Eagle r1 QPU \textit{ibm\_washington} (25 ansatz circuit tilings). Each VQE workload was submitted to the IBM Quantum service as a Qiskit Runtime job.

\hl{We have shared the code and data in a public GitHub repository \cite{N2_repo} so that the reader may reproduce the results of this work.}

\section{Results}\label{sec:results}
\begin{figure}
    \centering
    \includegraphics[width=0.8\linewidth]{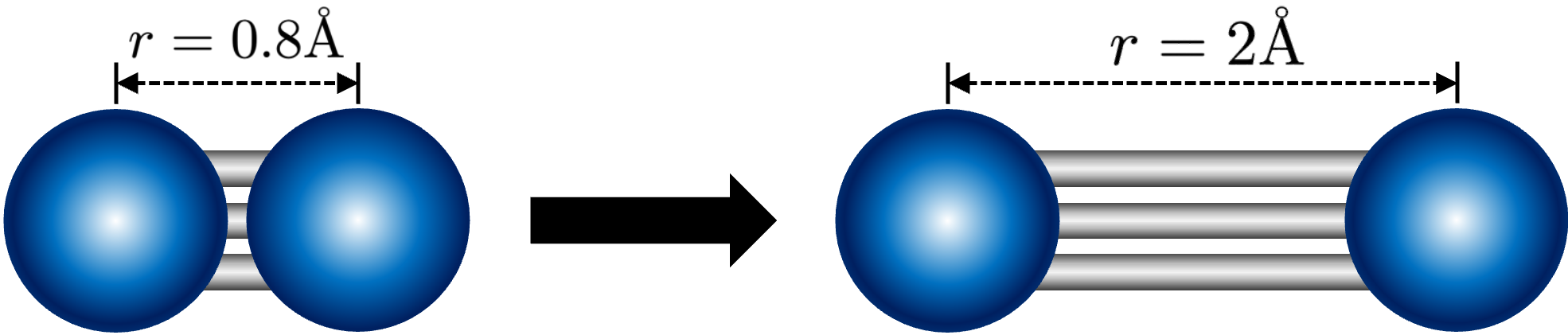}
    
    \vspace{3mm}
    
    \includegraphics[width=\linewidth]{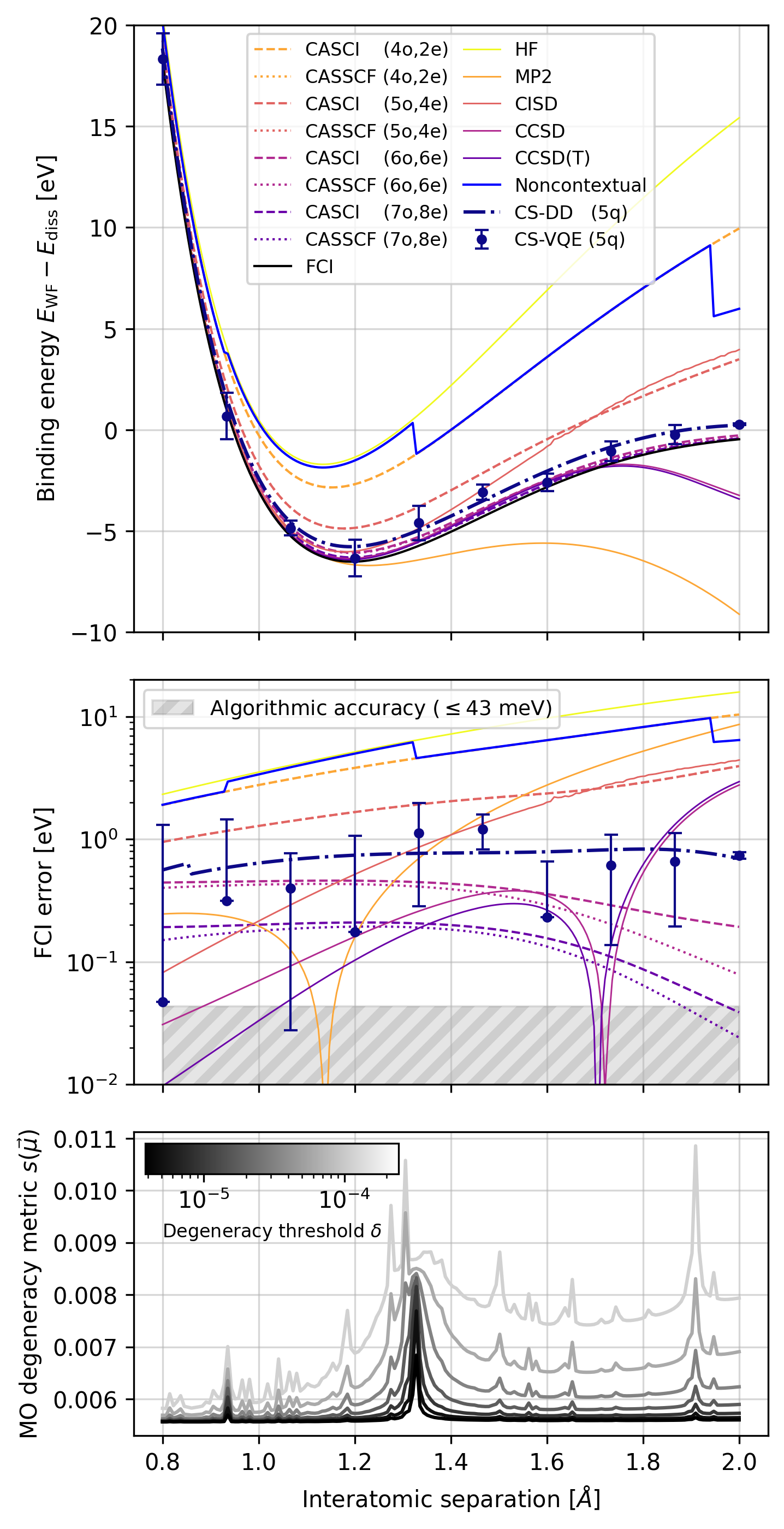}
    \caption{Binding potential energy curve for molecular nitrogen, \ce{N2}. \hl{The CS-VQE data points were evaluated on IBM Quantum hardware, while CS-DD corresponds with direct diagonalization of the five-qubit contextual subspaces.} The quantum simulations maintain good agreement with the full configuration interaction energy throughout the entire range of interatomic separations, \hl{outperforming all the single-reference methods in the dissociation limit and remaining competitive with CASCI/CASSCF at a considerable saving of qubits.} Discontinuities in the noncontextual energy coincide with peaks in our molecular-orbital degeneracy detection function \eqref{MO_score}. \hl{Bars indicate standard error on the mean.}}
    \label{fig:final_PEC}
\end{figure}

We perform Contextual Subspace Variational Quantum Eigensolver (CS-VQE) experiments for ten points along the binding potential energy curve (PEC) of \ce{N2} STO-3G, evenly spaced between $0.8\angstrom-2\angstrom$. \hl{The results of this may be viewed in Figure \ref{fig:final_PEC}. Alongside our experimental results, we include the following classical benchmarks: ROHF, MP2, CISD, CCSD, CCSD(T), CASCI and CASSCF. The active spaces of the latter two were selected using MP2 natural orbitals for fairness, since this is comparable to how the contextual subspaces were chosen as described in Section \ref{sec:subspace}. We included active spaces of varying sizes, specifically (4o,2e), (5o,4e), (6o,6e) and (7o,8e) where ($M$o,$N$e) denotes $N$ electrons correlated in $M$ spatial orbitals. A crucial point to note when comparing our CS-VQE results to the CAS methods is that, for an active subspace of $M$ spatial orbitals, one needs $2M$ qubits to represent the problem on a quantum computer; therefore, our chosen active spaces range from $8$ to $14$ qubits in size, while the contextual subspace consists of just 5 qubits. This is important to bear in mind when interpreting the results.}

In Figure \ref{fig:final_PEC} we see the single-reference quantum chemistry techniques - ROHF, MP2, CISD, CCSD and CCSD(T) - struggling to capture the FCI energy for \ce{N2}. This holds especially true in the dissociation limit where there is no agreement between the different approaches. While the conventional techniques yield relatively low error around the equilibirum length at $1.192\angstrom$ (albeit not within the target algorithmic accuracy of $43$~meV), they incur large error at stretched bond lengths due to a failure of restricted open-shell Hartree-Fock to describe static correlation. \hl{Furthermore, we see instances of non-variationality, which becomes apparent at $1.140 \angstrom$ for MP2, $1.706 \angstrom$ for CCSD(T) and $1.728 \angstrom$ CCSD. For the CAS methods, we do not capture the bond breaking appropriately until the active space is expanded to (6o,6e) or (7o,8e), corresponding with 12 and 14 qubit subspaces.}

\hl{On the other hand, our 5-qubit CS-VQE hardware experiments produce mean errors between $47$~meV and $1.2$~eV throughout the evaluated interatomic separations and remain below $1$~eV for all but two of the bond lengths ($1.333 \angstrom$ and $1.467 \angstrom$). From direct diagonalization (the CS-DD curve in Figure \ref{fig:final_PEC}) we see a true error range of $0.5$~eV to $0.8$~eV along the \ce{N2} PEC. Our quantum simulations outperform all the single-reference techniques in capturing the bond dissociation behaviour; indeed, beyond an interatomic separation of $1.351 \angstrom$ the 5-qubit contextual subspace calculation yields lower errors than CISD and MP2, and outperforms CCSD(T) after $1.834 \angstrom$, the `gold standard' of quantum chemistry.}

\hl{Turning now to the multiconfigurational approaches, our CS-VQE experiments produce lower error than CASCI/CASSCF (4o,2e) and (5o,4e) for every bond length, despite them corresponding with 8- and 10-qubit subspaces. In order for CAS to capture the dissociation satisfactorily, it needs at least the (6o,6e) space to describe the triple bond between nitrogen atoms appropriately; this is precisely what we find in Figure \ref{fig:final_PEC}. While the (6o,6e) and (7o,8e) calculations do yield improved errors, particularly towards the dissociation limit, we stress that they correspond with 12- and 14-qubit subspaces. Contextual subspaces of the same size yield considerably lower error than CAS in this instance (assessed through direct diagonalization), albeit caveated with the added challenge of running hardware experiments of that scale, which could prohibit us from achieving this in practice.}

Our energy advantage in the dissociation limit can be attributed to the noncontextual energy component of the CS-VQE simulations. Around the equilibrium length there is negligible difference between the Hartree-Fock and noncontextual energy, but as the bond is stretched we see the noncontextual approximation outperforming Hartree-Fock, even before the inclusion of contextual corrections obtained from VQE simulations. \hl{Interestingly, the noncontextual energy coincides exactly with the CASCI/CASSCF (4o,2e) curve between bond lengths $1.328 \angstrom$ and $1.909 \angstrom$.} This is because the noncontextual problem can accommodate a ground state that is multireference in nature, thus capturing the separated atom limit more appropriately than the single-reference ROHF state here. We note that the CS-VQE optimization is still initialized in the Hartree-Fock state and therefore does not receive an unfair advantage from this feature of the method; instead, the noncontextual contribution is included in the construction of the contextual Hamiltonian as a constant shift.

Curiously, the noncontextual PEC is not continuous and these error improvements are encountered in sharp decreases of energy, as seen in Figure \ref{fig:final_PEC} for interatomic separations $0.936 \angstrom, 1.328 \angstrom$ and $1.909 \angstrom$. In order to probe this effect, we search for degeneracy in the energy levels between molecular orbitals (MO), which is known to cause issues for MP \cite{lee2018regularized, hollett2020capturing}. This is achieved by detecting near-zero energy differences between elements of $\bm{\mu}$, a vector with length the number of orbitals $M$ whose entries are the canonical MO energies computed through Hartree-Fock. Our candidate MO degeneracy detection function is
\begin{equation}\label{MO_score}
\begin{aligned}
    s_{\delta}(\bm{\mu}) \coloneqq{} & \frac{\delta\sqrt{\pi}}{2(D_{\mathrm{max}}-1)} \sum_{i=2}^{D_{\mathrm{max}}} \frac{1}{{M \choose i}} \Bigg[ \\
    & \hspace{3mm} \sum_{1 \leq j_1 < \dots < j_i \leq M} \frac{\mathrm{erf}\big[(\mu_{j_i} -  \sum_{k=1}^{i}\mu_{j_k})/\delta\big]}{(\mu_{j_i} -  \sum_{k=1}^{i}\mu_{j_k})} \Bigg]
\end{aligned}
\end{equation}
where $\delta\geq0$ acts as a filtering parameter determining the threshold of near-degeneracy between energy levels, noting $\lim_{\delta \rightarrow 0} \frac{\sqrt{\pi}\delta\mathrm{erf}(x/\delta)}{2x} = \delta_{x,0}$ and thus for $\delta=0$ this will detect exact degeneracy. \hl{The metric satisfies $0 \leq s_{\delta}(\bm{\mu}) \leq 1$ and for $\delta < \delta^\prime$ we have $s_{\delta}(\bm{\mu}) \leq s_{\delta^\prime}(\bm{\mu})$.} The maximum depth $D_{\mathrm{max}} \leq M$ allows one to truncate the outer sum since the number of inner terms increases as ${M-1 \choose i}$; in the \ce{N2} STO-3G case $M = 10$ so we may include all MO degeneracy contributions, but for larger systems we may truncate for ease of computation. This may be viewed in the lower subplot of Figure \ref{fig:final_PEC}, where peaks indicate the presence of degenerate MOs. Encouragingly, these peaks coincide exactly with discontinuities in the noncontextual energy approximation, thus giving us confidence that the success of our Contextual Subspace techniques stems from its ability to describe static correlation in the noncontextual component. \hl{In Appendix \ref{sec:diagnostics} we supplement this new diagnostic tool with traditional approaches to detecting static correlation in wavefunction methods such as $T_1, D_1$ and correlation entropy, which may be viewed in Figure \ref{fig:diagnostics}.}
Finally, while we identified the occurrence of noncontextual discontinuities to coincide with peaks in our MO degeneracy metric \eqref{MO_score}, one might consider the converse implication of our noncontextual problem as itself a test for detecting non-dynamical correlation. This is closely related to the Coulson-Fischer point, characterized by a divergence between restricted and unrestricted Hartree-Fock calculations \cite{burton2021hartree}, indicating a break-down of spin symmetry.
\section{Conclusion}\label{sec:Conclusion}
\hl{This study unifies many contemporary techniques of quantum computation to deliver experimental results that challenge conventional quantum chemistry methods of a similar computational overhead.} In particular, it demonstrates the effectiveness of the Contextual Subspace method in surpassing classical single-reference techniques \hl{(ROHF, MP2, CISD, CCSD, CCSD(T)) and its competitiveness with more computationally-demanding multiconfigurational approaches (CASCI, CASSCF)}. While we do not claim to realize any form of quantum advantage (due to the problem being solvable at the FCI level), this work exemplifies the ability for our simulation methodology to capture static correlation effects (encountered in systems undergoing bond dissociation, for example). This stems from the noncontextual ground state being able to describe a superposition of electronic configurations, as opposed to a single Slater determinant. This is analogous to how multiconfigurational methods such as CASCI and CASSCF are able to treat non-dynamical correlations; \hl{however, we showed that a significantly reduced contextual subspace could describe the bond dissociation of \ce{N2} at a level comparable to active spaces more than twice the size in CAS methods.} A benefit of the Contextual Subspace technique is the active space is not explicitly selected \cite{stein2016automated, khedkar2019active} and the multi-reference character is informed by the choice of noncontextual Hamiltonian.

Future work will look into increasing the size of problem studied by expanding the basis set and/or molecular system. As we move to study systems outside the realm of FCI, \hl{we wish to assess whether our methodology is able to maintain its competitive advantage of significant reduction in quantum resource requirements, while still facilitating a similar level of accuracy as classical multiconfigurational techniques.} Furthermore, it is also possible to make a frozen core and/or active space approximation within the Contextual Subspace approach, a further consideration in which it will be interesting to explore the efficacy of this method. The Contextual Subspace methodology describes a flexible approach to quantum simulations that is scalable to large molecular systems. This should provide a practical route to quantum advantage and thus provide answers to scientifically-meaningful questions in the chemical domain.

\hl{\section{Data Availability}\label{data_availability}}
\hl{We provide all the data and code necessary to reproduce the results of this paper in a public GitHub repository \cite{N2_repo}.}

\section*{acknowledgements}
T.W. and A.R. acknowledge support from the Unitary Fund. T.W. also acknowledges support from EPSRC (EP/S021582/1), CBKSciCon Ltd., Atos, Intel and Zapata. A.R. and P.J.L. acknowledges support  by the NSF STAQ project (PHY-1818914). S.S. wishes to acknowledge financial support from the National Centre for HPC, Big Data and Quantum Computing (Spoke 10, CN00000013). P.V.C. is grateful for funding from the European Commission for VECMA (800925) and EPSRC for SEAVEA (EP/W007711/1). Access to the IBM Quantum Computers was obtained through the IBM Quantum Hub at CERN with which the Italian Institute of Technology (IIT) is affiliated.

\vspace{2cm}

\bibliographystyle{apsrev4-2_mod.bst}
\bibliography{main}

\clearpage
\onecolumngrid
\appendix

\begingroup
\renewcommand{\arraystretch}{1.5} 
\begin{table}
\section{VQE Experiment History}
\vspace{1.5cm}
\resizebox{\linewidth}{!}{
\small
\begin{tabular}{lllllll}
\hline
Year & Reference          & System(s)                          & Ansatz               & Max qubits & Platform                      & Hardware Vendor \\ \hline
2013 & Peruzzo  \textit{et al.} \cite{peruzzo2014variational}    & \ce{HeH+}                            & UCC          & 2          & SP              & In-house         \\
2015 & Shen  \textit{et al.} \cite{Shen2017}       & \ce{HeH+}                            & UCC          & 1 qudit          & TI                   & In-house                \\
2015 & Google Quantum \cite{OMalleyBabbush2016}     & \ce{H2}                              & UCC          & 2          & SC               & Google          \\
2016 & Santagati \textit{et al.} \cite{Santagati2018a}    & Chlorophyll pair                 &  Parametrized Hamiltonian    & 2          & SP              & In-house                \\
2017 & Kandala \textit{et al.} \cite{Kandala2017}                & \ce{H2},  \ce{LiH},  \ce{BeH2} & Hardware Efficient                  & 6          & SC               & IBM             \\
2017 & Colless \textit{et al.}\cite{Colless2018}           & \ce{H2} (excited states)             & Hardware Efficient                  & 2          & SC               & In-house       \\
2018 & Hempel \textit{et al. } \cite{hempel2018quantum}     & \ce{H2},  \ce{LiH}                         & UCC           & 3          & TI                   &     In-house            \\
2018 & Kandala \textit{et al.} \cite{Kandala2019}                & \ce{H2},  \ce{LiH} (magnetism)             & Hardware Efficient                  & 4          & SC               &    IBM         \\
2019 & Nam \textit{et al.} \cite{Nam2020}         & \ce{H2O}                             & UCC           & 4          & TI                   &  IonQ               \\
2019 & Smart \& Mazziotti \cite{Smart2019} & \ce{H3}                              & custom          & 3          & SC               &     IBM            \\
2019 & McCaskey \textit{et al.} \cite{McCaskey2019}         & \ce{NaH},  \ce{RbH},  \ce{KH}                    & UCC and Hardware Efficient          & 4          & SC               &    IBM, Rigetti             \\
2020 & Rice \textit{et al. } \cite{Rice2021}       & \ce{LiH} (dipole moment)             & Hardware Efficient                  & 4          &   SC                & IBM             \\
2020 & Google AI Quantum \cite{arute2020hartree}             & \ce{H6},  \ce{H8},  \ce{H10}, \ce{H12}, \ce{HNNH}           & Hartree-Fock         & 12         & SC               &     Google         \\
2020 & Gao \textit{et al.} \cite{Gao2021a}                & PSPCz      & $R_{y}$             & 2          & SC               &    IBM             \\
2021 & Kawashima \textit{et al. } \cite{Kawashima2021}  & \ce{H10}                             & qubit-CC             & 2          &         TI                      & IonQ            \\
2021 & Eddins\textit{et al.} \cite{eddins2022doubling}      & \ce{H2O}                             & Entanglement Forging & 5          &     SC            & IBM             \\
2021 & Yamamoto \textit{et al. } \cite{Yamamoto2022}   & Crystalline Iron Model          & UCCSD-PBC            & 2          &  SC        & IBM             \\
2021 & Kirsopp \textit{et al.  } \cite{Kirsopp2022}   & Oxazine derivatives             & YXXX                 & 4          & SC, TI & IBM, Quantinuum \\
2022 & Huang \textit{et al. } \cite{huang2022variational}    & \ce{H2}, \ce{CO}                & Linear Response                & 4         &    SC  & In-house          \\
2022 & Lolur \textit{et al.} \cite{lolur2023reference}      & \ce{HeH+}, \ce{LiH}                           & Hardware Efficient             & 4         &     SC  & IBM            \\
2022 & Leyton-Ortega \textit{et al. } \cite{leyton2022quantum}       & \ce{H2}              & UCCSD                & 4          &  SC             & IBM      \\
2022 & Liang \textit{et al.} \cite{liang2023napa}      & \ce{H2}, \ce{HeH+}, \ce{LiH}, \ce{H2O}, \ce{NaH}, \ce{CO2}                          & NAPA             & 6         &     SC  & IBM            \\
2022 & Motta \textit{et al. } \cite{Motta2022}      & \ce{H3S+}                            & Entanglement Forging & 6          &     SC          & IBM             \\
2022 & O'Brien \textit{et al. } \cite{OBrien2022}    & Cyclobutene Ring                & upCCD                & 10         &    SC  & Google          \\
2022 & Khan \textit{et al.} \cite{khan2023chemically}       & \ce{CH4}              & UCCSD                & 6          &  TI             & Quantinuum      \\

2022 & Zhao \textit{et al.} \cite{zhao2023orbital}      & \ce{Li2O}                            & oo-upCCD             & 12         &     TI   & IonQ \\

2022 & Guo \textit{et al.} \cite{guo2023experimental}      & \ce{H2}, \ce{LiH}, \ce{F2}                            & UCCSD             & 12         &     SC  & Zuchongzhi 2.0            \\
2023 & Weaving \textit{et al.} \cite{weaving2023benchmarking}      & \ce{HCl}                            & Hardware Efficient             & 3         &     SC  & IBM            \\
2023 & Liu \textit{et al.} \cite{liu2023performing}      & \ce{H2}, \ce{HeH+}                            & Hardware Efficient             & 1 qudit         &     SC  & In-house            \\
2023 & Dimitrov \textit{et al.} \cite{dimitrov2023pushing} & \ce{CH3F} & pUCCD & 11 & TI & IonQ \\
2023 & Jones \textit{et al.} \cite{jones2023precision}      & \ce{H2O}                            & UCCD             & 8         &     SC  & IBM            \\
2023 & Liang \textit{et al.} \cite{liang2023spacepulse}      & \ce{NH}, \ce{BeH+}, \ce{F2}  & SpacePulse             & 6         &     SC  & IBM            \\
2023 & Weaving \textit{et al.} [This work]      & \ce{N2}                            & Hardware-Aware ADAPT             & 5         &     SC  & IBM            \\ \hline
\end{tabular}
}
\caption{A decade of experimental realizations of VQE for quantum chemistry; the list is not exhaustive. The works are listed chronologically by the date of initial preprint availability, not the final publication date. The platform keys are silicon photonic (SP), superconducting (SC) and trapped-ion (TI).}
\label{tab:VQE_exp_todate}
\end{table}
\endgroup
\clearpage
\section{Hardware-Aware ADAPT-VQE Algorithms}\label{sec:adapt_algos}

\hl{Here, we present the two algorithms used in constructing our ansatz circuits. First, Algorithm \ref{adapt_alg} describes the general ADAPT-VQE framework for a arbitrary excitation scoring function $f$, while our approach to incorporating hardware-awareness into $f$ is described in Algorithm \ref{topology_aware_bias}.}

\begin{algorithm}[b]
    \caption{qubit-ADAPT-VQE; our hardware-aware scoring function $f$ \eqref{new_score} is described in Algorithm \ref{topology_aware_bias}.}\label{adapt_alg}
    
    \SetKwInOut{Input}{Input}
    \SetKwInOut{Output}{Output}

    \Input{Operator pool $\mathcal{P}$, initial state $\ket{\psi_0}$, scoring function $f: \mathcal{P} \mapsto \mathbb{R}$, score tolerance $\delta_f>0$, convergence threshold $\delta_c>0$ and maximum number of iterations $n_{\mathrm{max}} \in \mathbb{N}$.}
    \Output{Optimized energy $E(\bm{\theta}) = \bra{\psi(\bm{\theta})} H \ket{\psi(\bm{\theta})}$ and ansatz $\ket{\psi(\bm{\theta})} = \prod_n e^{i \theta_n P_n} \ket{\psi_0}$.}
    
    $n \leftarrow 0$ \;
    $E_0 \leftarrow 0$ \;
    
    \While{$(\Delta_{f} > \delta_{f}) \land (\Delta_{c} > \delta_{c}) \land (n_{\mathrm{max}} > n)$}{
        Identify optimal pool operator:
        \begin{equation*}
            P_{n+1} \leftarrow \argmax_{P \in \mathcal{P}} |f(P; \psi_n)|,\;\;\; \Delta_f \leftarrow |f(P_{n+1}; \psi_n)|\;
        \end{equation*}
        Append term to growing ansatz:
        $$\ket{\psi_{n+1}} \leftarrow e^{i \theta_{n+1} P_{n+1}} \ket{\psi_n}\;$$
        
        Optimize parameters through VQE:
        \begin{equation*}
            \bm{\theta}_{n+1} leftarrow \argmin_{\bm{\theta} \in \mathbb{R}^{\times (n+1)}} E(\bm{\theta}),\;\;\; 
            E_{n+1} \leftarrow E(\bm{\theta}_{n+1}),\;\;\;
            \Delta_{c} \leftarrow |E_{n+1} - E_{n}|\;
        \end{equation*}
        
        $n \leftarrow n+1$ \;
        
    }

\end{algorithm}

\begin{algorithm}[b!]
    \caption{Hardware-aware biasing function evaluation}\label{topology_aware_bias}
    
    \SetKwInOut{Input}{Input}
    \SetKwInOut{Output}{Output}
    
    \Input{Pool operator $P \in \mathcal{P}$, optimal state $\ket{\psi}$ from previous ADAPT iteration, target topology graph $\mathcal{G}_{\mathrm{target}} = (\mathcal{N}_{\mathrm{target}}, \mathcal{E}_{\mathrm{target}})$, bias $b > 0$ and maximum search depth $D \in \mathbb{N}$.}
    \Output{$f(P)$, a score for the pool operator $P$.}
    \vspace{3mm}
    For two graphs $\mathcal{G}, \mathcal{H}$ the function $\mathrm{VF2^{++}}(\mathcal{G}, \mathcal{H})$ returns \textbf{True} if $\mathcal{G}$ is subgraph isomorphic to $\mathcal{H}$ and \textbf{False} otherwise\;
    Build the weighted graph $\mathcal{G}_{\mathrm{circuit}} = (\mathcal{N}_{\mathrm{circuit}}, \mathcal{E}_{\mathrm{circuit}})$ for $e^{i \theta P} \ket{\psi}$. Here, $|\mathcal{N}_{\mathrm{circuit}}|<|\mathcal{N}_{\mathrm{target}}|$ represent the circuit qubits and $\mathcal{E}_{\mathrm{circuit}} \subset \mathcal{N}_{\mathrm{circuit}}^{\times 2} \times \mathbb{N}$ indicate the presence of a nonlocal operation in-circuit, weighted by the total number of occurrences. The sum of weights is $W = \sum_{(u,v,w) \in \mathcal{E}_{\mathrm{circuit}}} w.$
    
    $\Delta_f \leftarrow \frac{\partial}{\partial \theta} \bra{\psi} e^{- i \theta P} H e^{i \theta P} \ket{\psi} \big|_{\theta = 0}$, `standard' score \eqref{standard_ADAPT_score} \;
    $d \leftarrow 0$, the subgraph isomorphism distance \;
    \eIf{
        $\mathrm{VF2^{++}}(\mathcal{G}_{\mathrm{circuit}}, \mathcal{G}_{\mathrm{target}})$
        }{
            Already subgraph isomorphic -- no biasing\;
            \Return{$\Delta_f$}
        }{
    \While{$D > d$}{
        $d \leftarrow d+1$ \;
        Order node collections $\bm{n} \in \mathcal{N}_{\mathrm{circuit}}^{\times d}$ of size $d$ by their summed edge-weights $
            s(\bm{n}) \coloneqq \sum_{n \in \bm{n}} \sum_{\substack{(u,v,w) \in \mathcal{E}_{\mathrm{circuit}} \\ n = u \;\text{or}\; n = v}} w;
        $
        
        \For{
            $\bm{n} \in \argsort_{\bm{n} \in \mathcal{N}_{\mathrm{circuit}}^{\times d}}{s(\bm{n})}$
            }{
            Form the subgraph $\mathcal{G}(\bm{n}) \subset \mathcal{G}_{\mathrm{circuit}}$ in which the nodes $\bm{n}$ have been deleted from $\mathcal{G}_{\mathrm{circuit}}$\;
            \If{
                $\mathrm{VF2^{++}}(\mathcal{G}(\bm{n}), \mathcal{G}_{\mathrm{target}})$
                }{
                \Return{$\Delta_f \cdot (1-s(\bm{n})/W)^{b}$}
                }
            }
        
        }
        If the maximum depth is reached without finding a subgraph isomorphism, the score is set to zero\;
        \Return{$0$}
    }
     
\end{algorithm}

\begin{figure}

    \centering
    \section{Ansatz Circuits}\label{sec:ansatz_circuits}
    \vspace{1cm}
    \begin{subfigure}{\linewidth}
    \begin{quantikz}[column sep=2mm, row sep=2mm]
    \lstick{$\ket{0}_0$} & \gate{\sqrt{X}} & \gate{R_z(\frac{\pi}{2})} & \ctrl{1} & \qw                  & \ctrl{1} & \gate{R_z(\theta_2)}      & \gate{\sqrt{X}} & \qw & \qw \\  
    \lstick{$\ket{0}_1$} & \gate{\sqrt{X}} & \gate{R_z(\frac{\pi}{2})} & \targ{}  & \gate{R_z(\theta_0)} & \targ{}  & \gate{R_z(\frac{\pi}{2})} & \gate{\sqrt{X}} & \gate{R_z(\frac{\pi}{2})} & \qw \\  
    \lstick{$\ket{0}_2$} & \gate{\sqrt{X}} & \gate{R_z(\frac{\pi}{2})} & \ctrl{2} & \qw                  & \ctrl{2} & \gate{R_z(\frac{\pi}{2})} & \gate{\sqrt{X}} & \gate{R_z(\frac{\pi}{2})} & \qw \\  
    \lstick{$\ket{0}_3$} & \gate{X}        & \qw                       & \qw      & \qw                  & \qw      & \qw                       & \qw             & \qw & \qw \\  
    \lstick{$\ket{0}_4$} & \gate{\sqrt{X}} & \gate{R_z(-\frac{\pi}{2})}& \targ{}  & \gate{R_z(\theta_1)} & \targ{}  & \gate{R_z(\frac{\pi}{2})} & \gate{\sqrt{X}} & \qw & \qw
    \end{quantikz}
    \caption{$x = 0.8\angstrom$; prepares $ \ket{\psi(\bm{\theta})} = e^{i\theta_2 Y_0} e^{i\theta_1 Y_4 X_3} e^{i\theta_0 X_1 Y_0} \ket{1_41_30_20_10_0}$} 
    \end{subfigure}
    
    
    \begin{subfigure}{\linewidth}
    \begin{quantikz}[column sep=2mm, row sep=2mm]
    \lstick{$\ket{0}_0$} & \gate{\sqrt{X}} & \gate{R_z(\frac{\pi}{2})} & \ctrl{1} & \qw     & \qw                  & \qw      & \ctrl{1} & \gate{R_z(\frac{\pi}{2})} & \gate{\sqrt{X}} & \qw & \qw \\  
    \lstick{$\ket{0}_1$} & \gate{\sqrt{X}} & \gate{R_z(\frac{\pi}{2})} & \targ{}  & \ctrl{3}& \qw                  & \ctrl{3} & \targ{}  & \gate{R_z(\frac{\pi}{2})} & \gate{\sqrt{X}} & \gate{R_z(\frac{\pi}{2})} & \qw \\  
    \lstick{$\ket{0}_2$} & \gate{\sqrt{X}} & \gate{R_z(\frac{\pi}{2})} & \ctrl{1} & \qw     & \qw                  & \qw      & \ctrl{1} & \gate{R_z(\frac{\pi}{2})} & \gate{\sqrt{X}} & \qw & \qw \\  
    \lstick{$\ket{0}_3$} & \gate{\sqrt{X}} & \gate{R_z(-\frac{\pi}{2})}& \targ{}  & \qw     & \gate{R_z(\theta_0)} & \qw      & \targ{}  & \gate{R_z(\frac{\pi}{2})} & \gate{\sqrt{X}} & \gate{R_z(\frac{\pi}{2})} & \qw \\  
    \lstick{$\ket{0}_4$} & \gate{\sqrt{X}} & \gate{R_z(-\frac{\pi}{2})}& \qw      & \targ{} & \gate{R_z(\theta_1)} & \targ{}  & \qw      & \gate{R_z(\frac{\pi}{2})} & \gate{\sqrt{X}} & \gate{R_z(\frac{\pi}{2})} & \qw
    \end{quantikz}
    \caption{$0.8 < x \leq 1.2$\angstrom; prepares $ \ket{\psi(\bm{\theta})} = e^{i\theta_1 Y_4 Y_1 Y_0} e^{i\theta_0 X_3 Y_2} \ket{1_41_30_20_10_0} $}
    \end{subfigure}
    
    
    \begin{subfigure}{\linewidth}
    \begin{quantikz}[column sep=2mm, row sep=2mm]
    \lstick{$\ket{0}_0$} & \gate{\sqrt{X}} & \qw                       & \qw      & \qw                  & \qw      & \qw             & \qw             & \qw             & \qw      & \qw      & \qw                  & \qw      & \qw          & \gate{R_z(\theta_3)} & \gate{\sqrt{X}} & \qw \\  
    \lstick{$\ket{0}_1$} & \gate{\sqrt{X}} & \gate{R_z(\frac{\pi}{2})} & \ctrl{1} & \qw                  & \ctrl{1} & \qw             & \qw             & \qw             & \qw      & \qw      & \qw                  & \qw      & \qw          & \gate{R_z(\theta_4)} & \gate{\sqrt{X}} & \qw \\  
    \lstick{$\ket{0}_2$} & \gate{\sqrt{X}} & \gate{R_z(\frac{\pi}{2})} & \targ{}  & \gate{R_z(\theta_0)} & \targ{}  & \gate{R_z(\pi)} & \gate{\sqrt{X}} & \gate{R_z(\pi)} & \ctrl{1} & \qw      & \qw                  & \qw      & \ctrl{1}     & \gate{R_z(\frac{\pi}{2})} & \gate{\sqrt{X}} & \qw \\  
    \lstick{$\ket{0}_3$} & \gate{\sqrt{X}} & \gate{R_z(\frac{\pi}{2})} & \ctrl{1} & \qw                  & \ctrl{1} & \qw             & \qw             & \qw             & \targ{}  & \ctrl{1} & \qw                  & \ctrl{1} & \targ{}      & \gate{R_z(\frac{\pi}{2})} & \gate{\sqrt{X}} & \qw \\  
    \lstick{$\ket{0}_4$} & \gate{\sqrt{X}} & \gate{R_z(-\frac{\pi}{2})}& \targ{}  & \gate{R_z(\theta_1)} & \targ{}  & \gate{R_z(\pi)} & \gate{\sqrt{X}} & \gate{R_z(\pi)} & \qw      & \targ{}  & \gate{R_z(\theta_2)} & \targ{}  & \qw     & \gate{R_z(\frac{\pi}{2})} & \gate{\sqrt{X}} & \qw
    \end{quantikz}
    \caption{$1.2 < x < 2.0\angstrom$; prepares $  \ket{\psi(\bm{\theta})} = e^{i\theta_4 Y_1} e^{i\theta_3 Y_0} e^{i\theta_2 Y_4 Y_3 Y_2} e^{i\theta_1 X_4 Y_3} e^{i\theta_0 X_2 Y_1} \ket{1_40_30_20_10_0}$}
    \end{subfigure}
    
    
    \begin{subfigure}{\linewidth}
    \begin{quantikz}[column sep=2mm, row sep=2mm]
    \lstick{$\ket{0}_0$} & \gate{\sqrt{X}} & \gate{R_z(\theta_0)}      & \gate{\sqrt{X}} & \ctrl{1} & \qw                  & \ctrl{1} & \qw      & \qw             &  \qw          & \gate{R_z(\frac{\pi}{2})} & \gate{\sqrt{X}} & \gate{R_z(\frac{\pi}{2})} & \qw \\  
    \lstick{$\ket{0}_1$} & \gate{\sqrt{X}} & \gate{R_z(\frac{\pi}{2})} & \qw             & \targ{}  & \gate{R_z(\theta_3)} & \targ{}  & \ctrl{3} & \qw             &  \ctrl{3}     & \gate{R_z(\frac{\pi}{2})} & \gate{\sqrt{X}} & \qw & \qw \\  
    \lstick{$\ket{0}_2$} & \gate{\sqrt{X}} & \gate{R_z(\theta_1)}      & \gate{\sqrt{X}} & \ctrl{1} & \qw                  & \ctrl{1} & \qw      & \qw             &  \qw          & \gate{R_z(\frac{\pi}{2})} & \gate{\sqrt{X}} & \gate{R_z(\frac{\pi}{2})} & \qw \\  
    \lstick{$\ket{0}_3$} & \gate{\sqrt{X}} & \gate{R_z(\frac{\pi}{2})} & \qw             & \targ{}  & \gate{R_z(\theta_4)} & \targ{}  & \qw      & \qw             &  \qw          & \gate{R_z(\frac{\pi}{2})} & \gate{\sqrt{X}} & \qw & \qw \\  
    \lstick{$\ket{0}_4$} & \gate{\sqrt{X}} & \gate{R_z(\theta_2)}      & \gate{\sqrt{X}} & \qw      & \qw                  & \qw      & \targ{}  & \gate{R_z(\theta_5)} &  \targ{} & \gate{R_z(\frac{\pi}{2})} & \gate{\sqrt{X}} & \gate{R_z(\frac{\pi}{2})} & \qw
    \end{quantikz}
    \caption{$x = 2.0\angstrom$; prepares $ \ket{\psi(\bm{\theta})} = e^{i \theta_5 Y_1 X_4} e^{i \theta_4 Y_3 X_2} e^{i \theta_3 Y_1 X_0} e^{i \theta_2 Y_4} e^{i \theta_1 Y_2} e^{i \theta_0 Y_0} \ket{0_40_30_20_10_0}$}
    \end{subfigure}

    \caption{Ansatz constructions for differing bond lengths $0.8 \leq x \leq 2.0\angstrom$ for the \ce{N2} simulation of Section \ref{sec:results}, expressed in the IBM Quantum native gate-set. Excitations were selected using Hardware-Aware ADAPT-VQE, described in Section \ref{sec:ansatz_construction}, and the resulting circuits were heavily optimized for compactness.}
    \label{fig:ansatz_constructions}
\end{figure}
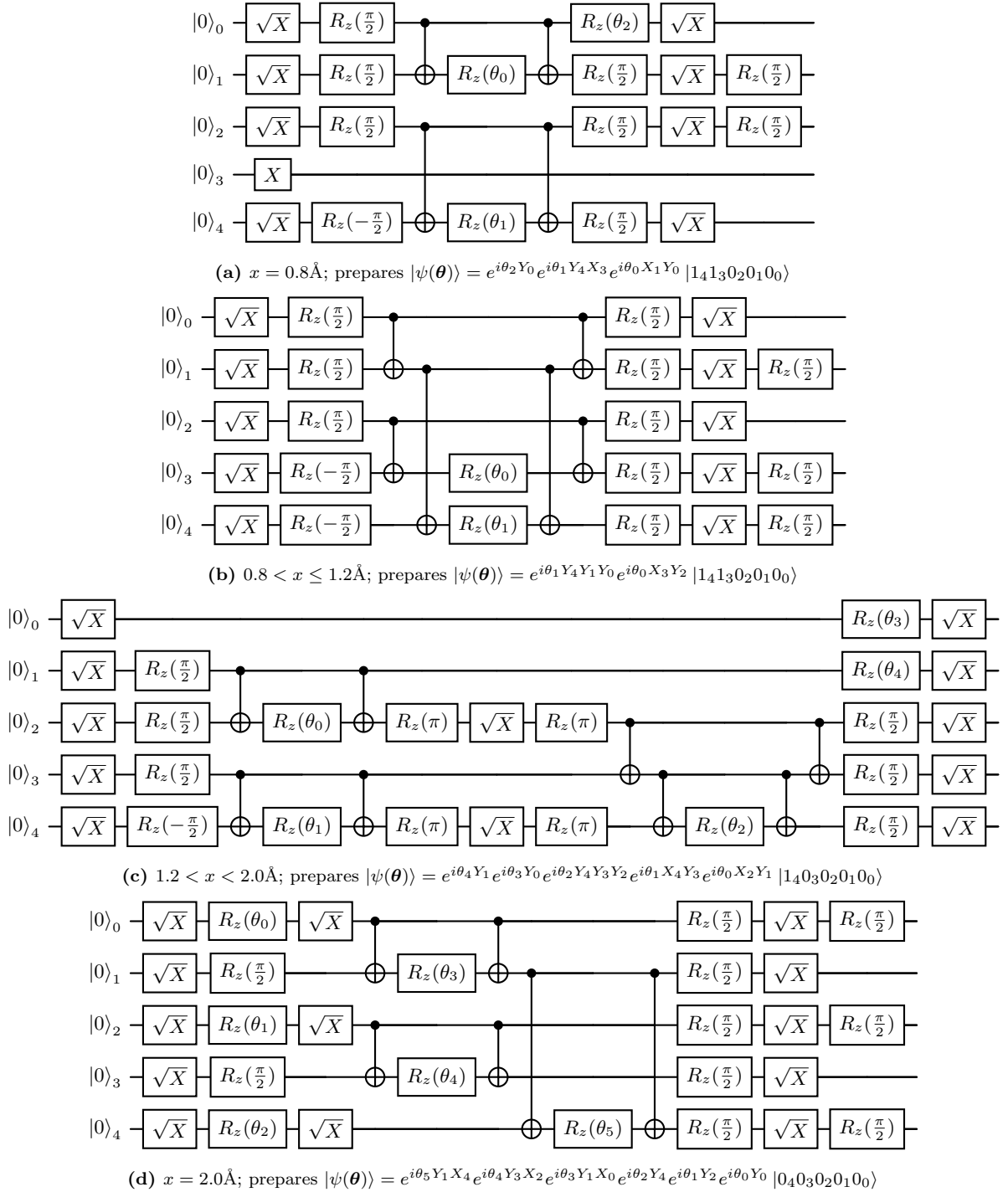

\clearpage
\hl{\section{Correlation Diagnostics}\label{sec:diagnostics}}

\hl{In Section \ref{sec:results} we introduced a new diagnostic tool for detecting (near) degeneracy between the energy levels of molecular orbitals. The function $s_{\delta}(\bm{\mu})$ is defined in Equation \eqref{MO_score} and provides a parameter $\delta$ that controls the degeneracy tolerance. We developed this metric to aid in probing the source of noncontextual discontinuities in Figure \ref{fig:final_PEC}, which we discovered coincide with peaks in $s_{\delta}(\bm{\mu})$.}

\begin{figure}[b]
    \centering
    \includegraphics[width=0.53\linewidth]{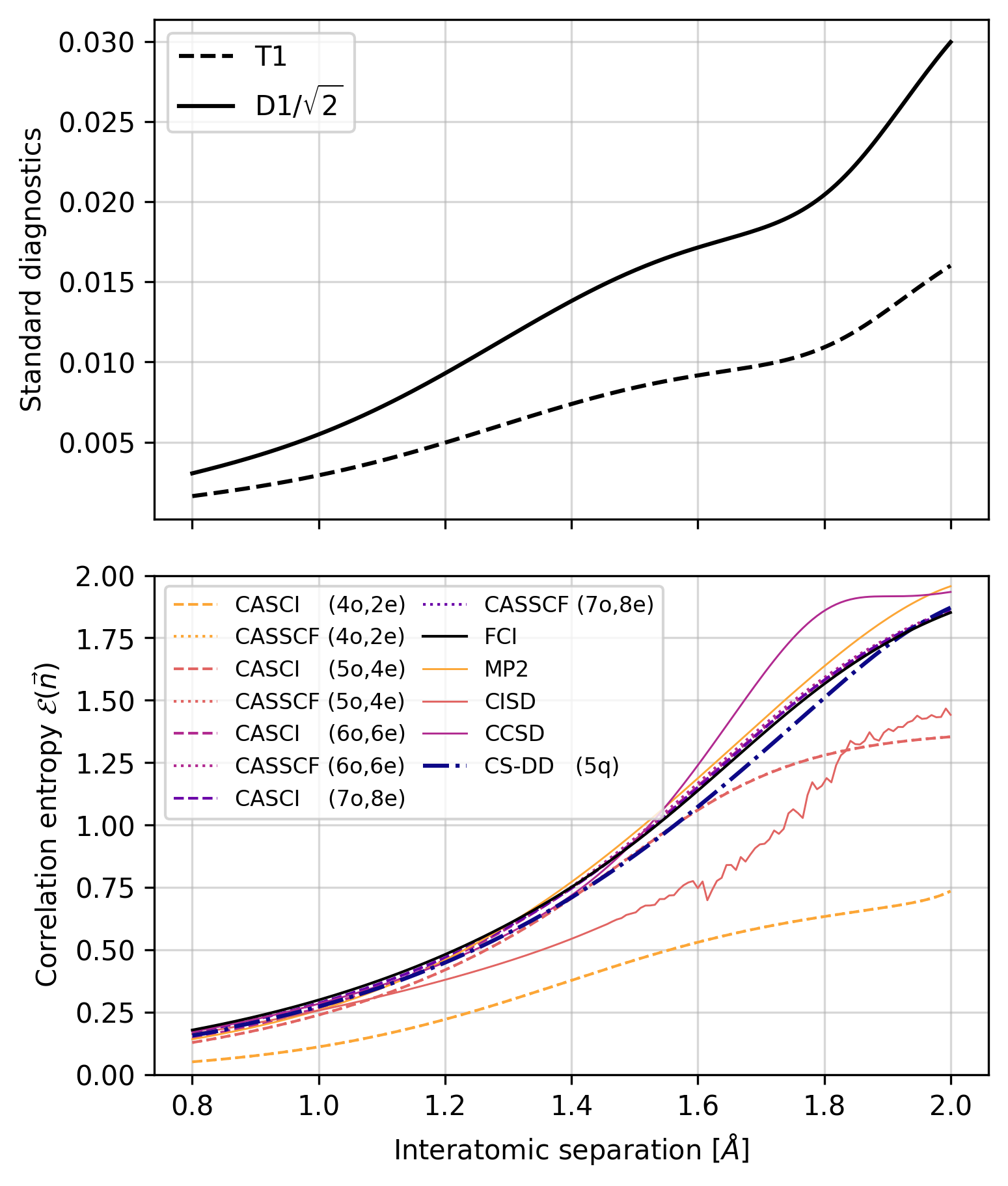}
    \caption{\hl{A collection of typical diagnostic tools for assessing the correlations present in various wavefunction techniques.}}
    \label{fig:diagnostics}
\end{figure}

In this section we supplement our new diagnostic tool with traditional approaches to detecting static correlation. For example, with $\bm{t}_1$ the vector of single-excitation amplitudes obtained from a coupled-cluster calculation, the $T_1$ diagnostic is defined as $||\bm{t}_1||_2/\sqrt{N_{\mathrm{elec}}}$ \cite{lee1989diagnostic}. It is often assumed that CCSD is reliable for $T_1<0.02$; however, in Figure \ref{fig:diagnostics} we see that $T_1$ never exceeds this threshold, even when CCSD becomes non-variational at $1.727 \angstrom$, and thus it is not a fool-proof metric. A related diagnostic is $D_1$, calculated as the largest singular value of the single-excitation amplitude matrix \cite{janssen1998new} and satisfies the inequality $D_1 \geq \sqrt{2}\cdot T_1$. Correlation entropy is another quantity that can provide an indication of strong non-dynamical correlation, which is defined as the Von Neumann entropy of the natural orbital occupation numbers (NOON) $\bm{n}$,
\begin{equation}
    \mathcal{E}(\bm{n}) = - \sum_i \frac{n_i}{2} \log{\Big(\frac{n_i}{2}\Big)},
\end{equation}
where $0\leq n_i \leq2$ represents the average number of particles filling orbital $i$. In Figure \ref{fig:diagnostics} we calculate the correlation entropy with NOONs obtained from the 5-qubit contextual subspace, CCSD and FCI (exact) calculations. Above an interatomic separation of $1.494 \angstrom$ the CCSD-derived NOONs over-approximate the entropy and, interestingly, the entropy plateaus where the non-variational character becomes apparent. On the other hand, the NOONs obtained from the 5-qubit contextual subspace follow more closely the FCI correlation entropy. A review of these and alternative techniques can be found in
\cite{fogueri2013simple}.

\hl{\section{STO-3G versus STO-6G}}

\begin{figure}[b]
    \centering
    \begin{minipage}{.5\textwidth}
        \centering
    \includegraphics[width=0.9\linewidth]{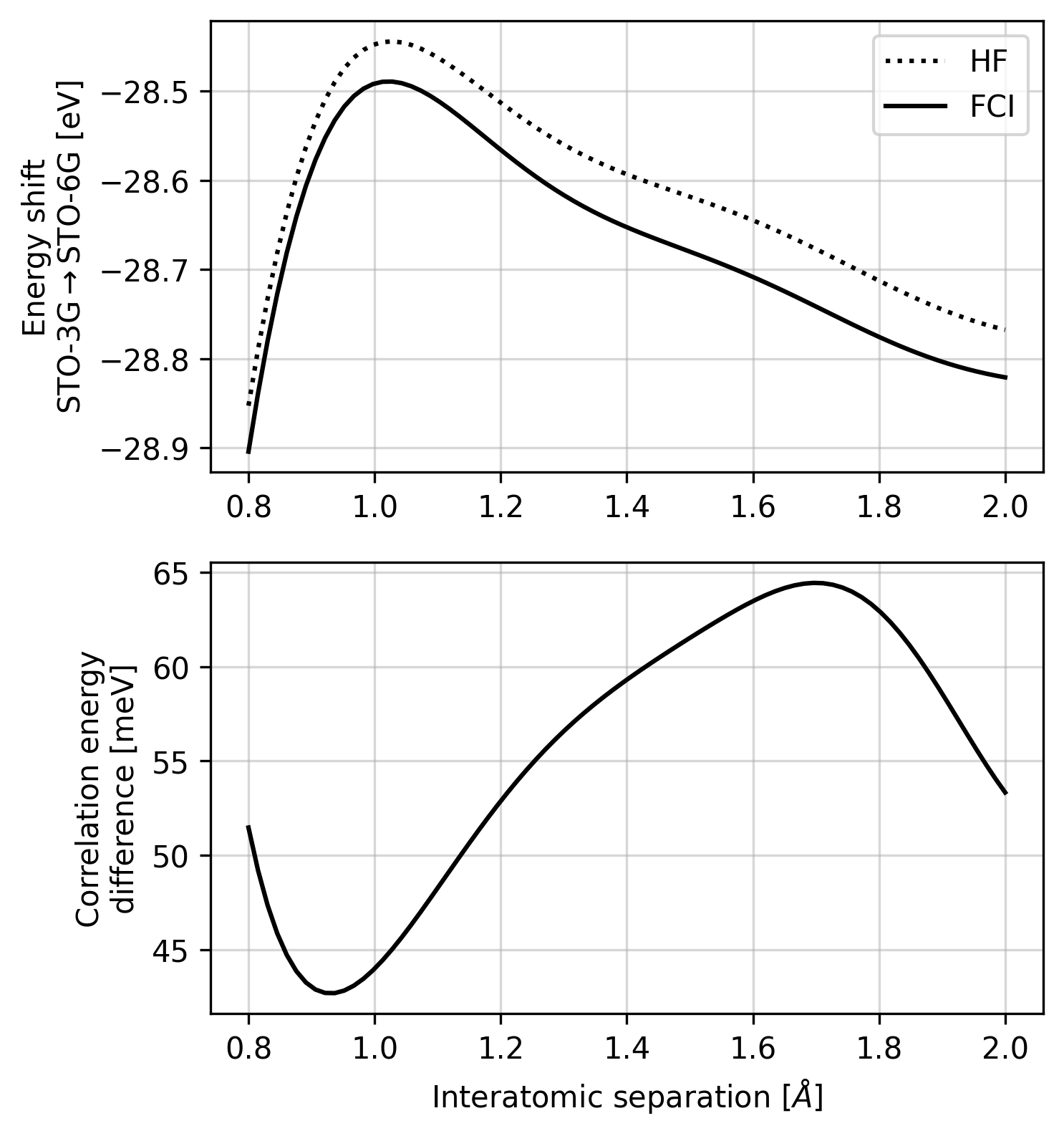}
    \caption{\hl{Comparison of ROHF and FCI between minimal basis sets STO-3G and STO-6G. The difference in correlation energy, while small in relation to the absolute energy shift between the two basis sets, is above the target threshold of algorithmic accuracy (within $43$ meV of FCI) and therefore can not be neglected.}}
    \label{fig:3G_vs_6G}
    \end{minipage}%
    \begin{minipage}{0.5\textwidth}
        \centering
    \includegraphics[width=0.9\linewidth]{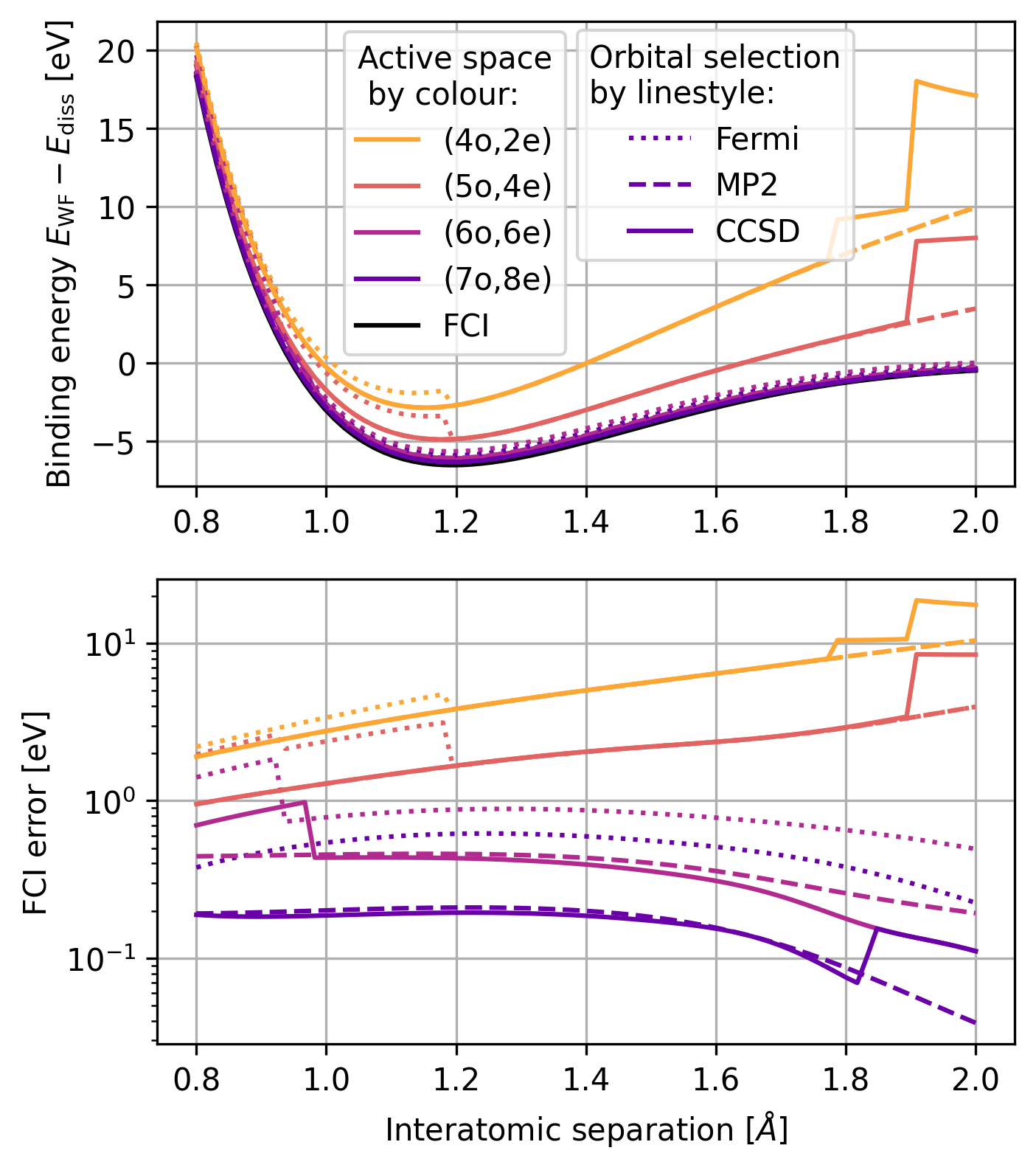}
    \caption{\hl{Comparison of active spaces in CASCI, both varying in size and the orbital selection criteria. The sizes investigated were (4o,2e),(5o,4e),(6o,6e) and (7o,8e), while the chosen orbitals were either selected around the Fermi level or from the MP2/CCSD natural orbitals.}}
    \label{fig:CASCI_disc}
    \end{minipage}
\end{figure}

\hl{The Slater-type orbital basis set STO-$n$G uses $n \in \mathbb{N}$ Gaussian functions to describe each atomic orbital. In the main text above we adopted the $n=3$ basis, however we compare here the difference between this and the $n=6$ variant. In Figure \ref{fig:3G_vs_6G}, going from $n=3$ to $n=6$ we see the FCI potential energy curve shifted down between $28.5-28.9$ eV across the bond lengths $0.8 - 2 \angstrom$. Despite this large energy shift, the difference in correlation energies $(E_{\mathrm{FCI}}^{(3G)} - E_{\mathrm{HF}}^{(3G)})-(E_{\mathrm{FCI}}^{(6G)} - E_{\mathrm{HF}}^{(6G)})$ only varies between $43 - 64$ meV; while this is small in comparison with the absolute shift, it still exceeds our target algorithmic accuracy $43$ meV and therefore cannot be disregarded. In future work we will consider the use of basis sets outside of STO-3G, particularly those tailored specifically for active space methods.}

\hl{\section{CASCI Active Space Selection Comparison}\label{sec:CASCI_active_space}}

\hl{In Section \ref{sec:subspace} we discussed how motivating contextual subspaces from either the MP2 or CCSD excitation generators would yield different subspaces. In Figure \ref{fig:CS_MP2_CCSD_comp} we found that, typically, the CCSD-motivated subspaces resulted in lower errors, but at the cost of discontinuities in the potential energy curve. As such, we argued that MP2-motivated spaces were preferable, favouring more stable dissociation curves over lower subspace error.}

\hl{For a fair comparison, we investigated how motivating the active space selection in CASCI via different means would affect the resulting energies. So that the methods are comparable, we used the MP2 and CCSD natural orbitals, in analogy with the contextual subspace motivation heuristic. Furthermore, we also tried simply selecting orbitals lying around the Fermi level, since naively this is where excitations are most energetically favourable.
}

\hl{The results of this may be viewed in Figure \ref{fig:CASCI_disc} and, surprisingly, we see a mirroring with the contextual subspace curves of Figure \ref{fig:CS_MP2_CCSD_comp}. Indeed, the Fermi- and CCSD-motivated active spaces exhibit discontinuities, while the MP2 dissociation does not. These results motivate our use of the MP2 natural orbitals throughout the work.}

\newpage

\begin{table} \section{Hamiltonians}\label{sec:hamiltonians}
 \caption{Qubit-wise commuting decomposition of the \ce{N2} molecular Hamiltonian at a separation of $ 0.80\angstrom$} \label{H0}      \begin{tabular}{|p{0.06\linewidth} | p{0.9\linewidth}|}     \hline     \textbf{Clique Index} & \textbf{QWC Hamiltonian Terms} \\     \hline
 Identity & $-103.58363 \cdot IIIII$ \\ \hline
 0 & $  + 0.14413 \cdot IIIIZ + 0.14413 \cdot IIIZI + 0.64710 \cdot IIIZZ + 0.14413 \cdot IIZII + 0.28439 \cdot IIZIZ + 0.29793 \cdot IIZZI + 0.14413 \cdot IIZZZ + 0.95822 \cdot IZIII + 0.26516 \cdot IZIIZ + 0.26516 \cdot IZIZI + 0.26516 \cdot IZZII - 0.27867 \cdot IZZIZ - 0.29314 \cdot IZZZI + 0.26516 \cdot IZZZZ + 0.56212 \cdot ZIIII + 0.28110 \cdot ZIIIZ + 0.29015 \cdot ZIIZI + 0.56212 \cdot ZIIZZ + 0.23447 \cdot ZIZII - 0.25837 \cdot ZIZIZ - 0.25837 \cdot ZIZZI + 0.31482 \cdot ZIZZZ + 0.25837 \cdot ZZIII - 0.28110 \cdot ZZIIZ - 0.29015 \cdot ZZIZI + 0.25837 \cdot ZZIZZ - 0.23447 \cdot ZZZII - 0.56212 \cdot ZZZIZ - 0.56212 \cdot ZZZZI - 0.31482 \cdot ZZZZZ $ \\ \hline1 & $ + 0.00905 \cdot IIXZZ + 0.01065 \cdot IXIII + 0.01065 \cdot IXIZZ + 0.02130 \cdot IXXII + 0.02130 \cdot IXXZZ + 0.01448 \cdot XIIZZ + 0.08035 \cdot XIXZZ + 0.01065 \cdot XXIII + 0.01065 \cdot XXIZZ $ \\ \hline2 & $ + 0.08035 \cdot IIIYY + 0.00452 \cdot IIYYI - 0.00452 \cdot IZYYI + 0.00905 \cdot XIIYY - 0.02661 \cdot XIYIY + 0.02661 \cdot XZYIY - 0.03113 \cdot XIYYI + 0.03113 \cdot XZYYI $ \\ \hline3 & $ - 0.02661 \cdot IIZXZ + 0.02661 \cdot IZZXZ + 0.02496 \cdot ZIIXZ + 0.00164 \cdot ZIZXI - 0.02496 \cdot ZZIXZ - 0.00164 \cdot ZZZXI $ \\ \hline4 & $ - 0.03113 \cdot IIZZX + 0.03113 \cdot IZZZX + 0.01448 \cdot XZZZI + 0.00452 \cdot XIZZX - 0.00452 \cdot XZZZX $ \\ \hline5 & $ + 0.03113 \cdot ZIIZX - 0.03113 \cdot ZZIZX + 0.00905 \cdot ZZXZI - 0.00617 \cdot ZIXZX + 0.00617 \cdot ZZXZX $ \\ \hline6 & $ + 0.02130 \cdot IYYZZ + 0.08035 \cdot YIYZZ + 0.01065 \cdot YYIZZ $ \\ \hline7 & $ + 0.08035 \cdot ZZZYY $ \\ \hline     \end{tabular}     \end{table}     

 \begin{table} \caption{Qubit-wise commuting decomposition of the \ce{N2} molecular Hamiltonian at a separation of $ 0.93\angstrom$} \label{H1}      \begin{tabular}{|p{0.06\linewidth} | p{0.9\linewidth}|}     \hline     \textbf{Clique Index} & \textbf{QWC Hamiltonian Terms} \\     \hline  
 Identity & $-104.60243 \cdot IIIII$ \\ \hline
 0 & $  + 0.17204 \cdot IIIIZ + 0.17204 \cdot IIIZI + 0.62346 \cdot IIIZZ + 0.17204 \cdot IIZII + 0.27535 \cdot IIZIZ + 0.28844 \cdot IIZZI + 0.17204 \cdot IIZZZ + 0.49343 \cdot IZIII + 0.27185 \cdot IZIIZ + 0.28110 \cdot IZIZI + 0.49343 \cdot IZIZZ + 0.21867 \cdot IZZII - 0.25019 \cdot IZZIZ - 0.25019 \cdot IZZZI + 0.30490 \cdot IZZZZ + 0.49343 \cdot ZIIII + 0.21867 \cdot ZIIIZ + 0.30490 \cdot ZIIZI + 0.49343 \cdot ZIIZZ + 0.27185 \cdot ZIZII - 0.25019 \cdot ZIZIZ - 0.25019 \cdot ZIZZI + 0.28110 \cdot ZIZZZ + 0.26913 \cdot ZZIII - 0.25643 \cdot ZZIIZ - 0.25643 \cdot ZZIZI + 0.28237 \cdot ZZIZZ - 0.25643 \cdot ZZZII - 0.96066 \cdot ZZZIZ - 0.25643 \cdot ZZZZZ $ \\ \hline1 & $ - 0.02079 \cdot IIXII - 0.02079 \cdot IIXZZ - 0.01146 \cdot IXIII - 0.01146 \cdot IXIZZ + 0.08623 \cdot IXXZZ - 0.01146 \cdot XIIII - 0.01146 \cdot XIIZZ + 0.00925 \cdot XIXZZ + 0.01324 \cdot XXIZZ $ \\ \hline2 & $ - 0.02079 \cdot IIIYY + 0.00925 \cdot IXIYY - 0.03386 \cdot IXYYI + 0.08623 \cdot XIIYY + 0.00463 \cdot XIYYI - 0.02923 \cdot XXYIY $ \\ \hline3 & $ - 0.00925 \cdot IYIYX + 0.03386 \cdot IYXYI - 0.08623 \cdot YIIYX - 0.00463 \cdot YIXYI - 0.02923 \cdot YYXIX $ \\ \hline4 & $ + 0.08623 \cdot IYYZZ + 0.00925 \cdot YIYZZ + 0.01324 \cdot YYIZZ $ \\ \hline     \end{tabular}     \end{table}     

 \begin{table} \caption{Qubit-wise commuting decomposition of the \ce{N2} molecular Hamiltonian at a separation of $ 1.07\angstrom$} \label{H2}      \begin{tabular}{|p{0.06\linewidth} | p{0.9\linewidth}|}     \hline     \textbf{Clique Index} & \textbf{QWC Hamiltonian Terms} \\     \hline  
 Identity & $-105.11963 \cdot IIIII$ \\ \hline
 0 & $  + 0.19316 \cdot IIIIZ + 0.19316 \cdot IIIZI + 0.60154 \cdot IIIZZ + 0.19316 \cdot IIZII + 0.26682 \cdot IIZIZ + 0.27946 \cdot IIZZI + 0.19316 \cdot IIZZZ + 0.43837 \cdot IZIII + 0.26322 \cdot IZIIZ + 0.27266 \cdot IZIZI + 0.43837 \cdot IZIZZ + 0.20340 \cdot IZZII - 0.24369 \cdot IZZIZ - 0.24369 \cdot IZZZI + 0.29571 \cdot IZZZZ + 0.43837 \cdot ZIIII + 0.20340 \cdot ZIIIZ + 0.29571 \cdot ZIIZI + 0.43837 \cdot ZIIZZ + 0.26322 \cdot ZIZII - 0.24369 \cdot ZIZIZ - 0.24369 \cdot ZIZZI + 0.27266 \cdot ZIZZZ + 0.26015 \cdot ZZIII - 0.24902 \cdot ZZIIZ - 0.24902 \cdot ZZIZI + 0.27237 \cdot ZZIZZ - 0.24902 \cdot ZZZII - 0.97165 \cdot ZZZIZ - 0.24902 \cdot ZZZZZ $ \\ \hline1 & $ - 0.02060 \cdot IIXII - 0.02060 \cdot IIXZZ - 0.01245 \cdot IXIII - 0.01245 \cdot IXIZZ + 0.09231 \cdot IXXZZ - 0.01245 \cdot XIIII - 0.01245 \cdot XIIZZ + 0.00945 \cdot XIXZZ + 0.01221 \cdot XXIZZ $ \\ \hline2 & $ - 0.02060 \cdot IIIYY + 0.00945 \cdot IXIYY - 0.03671 \cdot IXYYI + 0.09231 \cdot XIIYY + 0.00472 \cdot XIYYI - 0.03198 \cdot XXYIY $ \\ \hline3 & $ - 0.00945 \cdot IYIYX + 0.03671 \cdot IYXYI - 0.09231 \cdot YIIYX - 0.00472 \cdot YIXYI - 0.03198 \cdot YYXIX $ \\ \hline4 & $ + 0.09231 \cdot IYYZZ + 0.00945 \cdot YIYZZ + 0.01221 \cdot YYIZZ $ \\ \hline     \end{tabular}     \end{table}     

 \begin{table} \caption{Qubit-wise commuting decomposition of the \ce{N2} molecular Hamiltonian at a separation of $ 1.20\angstrom$} \label{H3}      \begin{tabular}{|p{0.06\linewidth} | p{0.9\linewidth}|}     \hline     \textbf{Clique Index} & \textbf{QWC Hamiltonian Terms} \\     \hline   
 Identity & $-105.37445 \cdot IIIII$ \\ \hline
 0 & $  + 0.20897 \cdot IIIIZ + 0.20897 \cdot IIIZI + 0.58184 \cdot IIIZZ + 0.20897 \cdot IIZII + 0.25885 \cdot IIZIZ + 0.27108 \cdot IIZZI + 0.20897 \cdot IIZZZ + 0.39475 \cdot IZIII + 0.25531 \cdot IZIIZ + 0.26493 \cdot IZIZI + 0.39475 \cdot IZIZZ + 0.18898 \cdot IZZII - 0.23849 \cdot IZZIZ - 0.23849 \cdot IZZZI + 0.28736 \cdot IZZZZ + 0.39475 \cdot ZIIII + 0.18898 \cdot ZIIIZ + 0.28736 \cdot ZIIZI + 0.39475 \cdot ZIIZZ + 0.25531 \cdot ZIZII - 0.23849 \cdot ZIZIZ - 0.23849 \cdot ZIZZI + 0.26493 \cdot ZIZZZ + 0.25201 \cdot ZZIII - 0.24272 \cdot ZZIIZ - 0.24272 \cdot ZZIZI + 0.26344 \cdot ZZIZZ - 0.24272 \cdot ZZZII - 0.98735 \cdot ZZZIZ - 0.24272 \cdot ZZZZZ $ \\ \hline1 & $ - 0.02060 \cdot IIXII - 0.02060 \cdot IIXZZ - 0.01353 \cdot IXIII - 0.01353 \cdot IXIZZ + 0.09837 \cdot IXXZZ - 0.01353 \cdot XIIII - 0.01353 \cdot XIIZZ + 0.00962 \cdot XIXZZ + 0.01143 \cdot XXIZZ $ \\ \hline2 & $ - 0.02060 \cdot IIIYY + 0.00962 \cdot IXIYY - 0.03957 \cdot IXYYI + 0.09837 \cdot XIIYY + 0.00481 \cdot XIYYI - 0.03476 \cdot XXYIY $ \\ \hline3 & $ - 0.00962 \cdot IYIYX + 0.03957 \cdot IYXYI - 0.09837 \cdot YIIYX - 0.00481 \cdot YIXYI - 0.03476 \cdot YYXIX $ \\ \hline4 & $ + 0.09837 \cdot IYYZZ + 0.00962 \cdot YIYZZ + 0.01143 \cdot YYIZZ $ \\ \hline     \end{tabular}     \end{table}     

 \begin{table} \caption{Qubit-wise commuting decomposition of the \ce{N2} molecular Hamiltonian at a separation of $ 1.33\angstrom$} \label{H4}      \begin{tabular}{|p{0.06\linewidth} | p{0.9\linewidth}|}     \hline     \textbf{Clique Index} & \textbf{QWC Hamiltonian Terms} \\     \hline  
 Identity & $-105.73143 \cdot IIIII$ \\ \hline
 0 & $ - 0.76333 \cdot IIIIZ - 0.04757 \cdot IIIZI + 0.03073 \cdot IIIZZ - 0.04757 \cdot IIZII + 0.03073 \cdot IIZIZ + 0.56455 \cdot IIZZI - 0.04757 \cdot IZIII + 0.03073 \cdot IZIIZ + 0.49632 \cdot IZIZI + 0.51907 \cdot IZZII - 0.04757 \cdot IZZZI + 0.03073 \cdot IZZZZ + 0.08875 \cdot ZIIII + 0.03744 \cdot ZIIIZ + 0.49635 \cdot ZIIZI + 0.51587 \cdot ZIZII + 0.08875 \cdot ZIZZI + 0.03744 \cdot ZIZZZ + 0.35124 \cdot ZZIII + 0.08875 \cdot ZZIZI + 0.03744 \cdot ZZIZZ + 0.08875 \cdot ZZZII + 0.03744 \cdot ZZZIZ + 0.55975 \cdot ZZZZI $ \\ \hline1 & $ - 0.12429 \cdot IIIIX - 0.10426 \cdot IIYYI - 0.00976 \cdot IYYII + 0.00976 \cdot XIYYI + 0.07498 \cdot XYIYI - 0.08474 \cdot XYYII $ \\ \hline2 & $ + 0.08474 \cdot IZZXI + 0.01088 \cdot XZZII + 0.00976 \cdot XZZXI $ \\ \hline3 & $ - 0.00976 \cdot IXZZI - 0.01088 \cdot XIZZI + 0.10426 \cdot XXZZI $ \\ \hline4 & $ - 0.08474 \cdot ZIZXI + 0.00976 \cdot ZXZII - 0.01097 \cdot ZXZXI $ \\ \hline5 & $ + 0.10426 \cdot ZZYYI $ \\ \hline6 & $ + 0.10426 \cdot YYZZI $ \\ \hline     \end{tabular}     \end{table}     

 \begin{table} \caption{Qubit-wise commuting decomposition of the \ce{N2} molecular Hamiltonian at a separation of $ 1.47\angstrom$} \label{H5}      \begin{tabular}{|p{0.06\linewidth} | p{0.9\linewidth}|}     \hline     \textbf{Clique Index} & \textbf{QWC Hamiltonian Terms} \\     \hline  
 Identity & $-105.94845 \cdot IIIII$ \\ \hline
 0 & $ - 0.60340 \cdot IIIIZ - 0.03004 \cdot IIIZI + 0.02599 \cdot IIIZZ - 0.03004 \cdot IIZII + 0.02599 \cdot IIZIZ + 0.54960 \cdot IIZZI - 0.03004 \cdot IZIII + 0.02599 \cdot IZIIZ + 0.48338 \cdot IZIZI + 0.50545 \cdot IZZII - 0.03004 \cdot IZZZI + 0.02599 \cdot IZZZZ + 0.07215 \cdot ZIIII + 0.03072 \cdot ZIIIZ + 0.48359 \cdot ZIIZI + 0.50335 \cdot ZIZII + 0.07215 \cdot ZIZZI + 0.03072 \cdot ZIZZZ + 0.32684 \cdot ZZIII + 0.07215 \cdot ZZIZI + 0.03072 \cdot ZZIZZ + 0.07215 \cdot ZZZII + 0.03072 \cdot ZZZIZ + 0.54648 \cdot ZZZZI $ \\ \hline1 & $ - 0.14665 \cdot IIIIX - 0.10982 \cdot IIYYI - 0.00988 \cdot IYYII + 0.00988 \cdot XIYYI + 0.08019 \cdot XYIYI - 0.09006 \cdot XYYII $ \\ \hline2 & $ + 0.09006 \cdot IZZXI + 0.01053 \cdot XZZII + 0.00988 \cdot XZZXI $ \\ \hline3 & $ - 0.00988 \cdot IXZZI - 0.01053 \cdot XIZZI + 0.10982 \cdot XXZZI $ \\ \hline4 & $ - 0.09006 \cdot ZIZXI + 0.00988 \cdot ZXZII - 0.01078 \cdot ZXZXI $ \\ \hline5 & $ + 0.10982 \cdot ZZYYI $ \\ \hline6 & $ + 0.10982 \cdot YYZZI $ \\ \hline     \end{tabular}     \end{table}     

 \begin{table} \caption{Qubit-wise commuting decomposition of the \ce{N2} molecular Hamiltonian at a separation of $ 1.60\angstrom$} \label{H6}      \begin{tabular}{|p{0.06\linewidth} | p{0.9\linewidth}|}     \hline     \textbf{Clique Index} & \textbf{QWC Hamiltonian Terms} \\     \hline  
  Identity & $-106.08444 \cdot IIIII$ \\ \hline
  0 & $ - 0.48091 \cdot IIIIZ - 0.01598 \cdot IIIZI + 0.02152 \cdot IIIZZ - 0.01598 \cdot IIZII + 0.02152 \cdot IIZIZ + 0.53673 \cdot IIZZI - 0.01598 \cdot IZIII + 0.02152 \cdot IZIIZ + 0.47193 \cdot IZIZI + 0.49353 \cdot IZZII - 0.01598 \cdot IZZZI + 0.02152 \cdot IZZZZ + 0.06019 \cdot ZIIII + 0.02477 \cdot ZIIIZ + 0.47224 \cdot ZIIZI + 0.49219 \cdot ZIZII + 0.06019 \cdot ZIZZI + 0.02477 \cdot ZIZZZ + 0.30479 \cdot ZZIII + 0.06019 \cdot ZZIZI + 0.02477 \cdot ZZIZZ + 0.06019 \cdot ZZZII + 0.02477 \cdot ZZZIZ + 0.53478 \cdot ZZZZI $ \\ \hline1 & $ - 0.16895 \cdot IIIIX - 0.11499 \cdot IIYYI - 0.00997 \cdot IYYII + 0.00997 \cdot XIYYI + 0.08507 \cdot XYIYI - 0.09504 \cdot XYYII $ \\ \hline2 & $ + 0.09504 \cdot IZZXI + 0.01032 \cdot XZZII + 0.00997 \cdot XZZXI $ \\ \hline3 & $ - 0.00997 \cdot IXZZI - 0.01032 \cdot XIZZI + 0.11499 \cdot XXZZI $ \\ \hline4 & $ - 0.09504 \cdot ZIZXI + 0.00997 \cdot ZXZII - 0.01065 \cdot ZXZXI $ \\ \hline5 & $ + 0.11499 \cdot ZZYYI $ \\ \hline     \end{tabular}     \end{table}     

 \begin{table} \caption{Qubit-wise commuting decomposition of the \ce{N2} molecular Hamiltonian at a separation of $ 1.73\angstrom$} \label{H7}      \begin{tabular}{|p{0.06\linewidth} | p{0.9\linewidth}|}     \hline     \textbf{Clique Index} & \textbf{QWC Hamiltonian Terms} \\     \hline    
 Identity & $-106.16893 \cdot IIIII$ \\ \hline
 0 & $ - 0.38444 \cdot IIIIZ - 0.00492 \cdot IIIZI + 0.01762 \cdot IIIZZ - 0.00492 \cdot IIZII + 0.01762 \cdot IIZIZ + 0.52564 \cdot IIZZI - 0.00492 \cdot IZIII + 0.01762 \cdot IZIIZ + 0.46182 \cdot IZIZI + 0.48309 \cdot IZZII - 0.00492 \cdot IZZZI + 0.01762 \cdot IZZZZ + 0.05151 \cdot ZIIII + 0.01982 \cdot ZIIIZ + 0.46217 \cdot ZIIZI + 0.48227 \cdot ZIZII + 0.05151 \cdot ZIZZI + 0.01982 \cdot ZIZZZ + 0.28499 \cdot ZZIII + 0.05151 \cdot ZZIZI + 0.01982 \cdot ZZIZZ + 0.05151 \cdot ZZZII + 0.01982 \cdot ZZZIZ + 0.52445 \cdot ZZZZI $ \\ \hline1 & $ - 0.18956 \cdot IIIIX - 0.11973 \cdot IIYYI - 0.01005 \cdot IYYII + 0.01005 \cdot XIYYI + 0.08959 \cdot XYIYI - 0.09964 \cdot XYYII $ \\ \hline2 & $ + 0.09964 \cdot IZZXI + 0.01020 \cdot XZZII + 0.01005 \cdot XZZXI $ \\ \hline3 & $ - 0.01005 \cdot IXZZI - 0.01020 \cdot XIZZI + 0.11973 \cdot XXZZI $ \\ \hline4 & $ - 0.09964 \cdot ZIZXI + 0.01005 \cdot ZXZII - 0.01055 \cdot ZXZXI $ \\ \hline5 & $ + 0.11973 \cdot ZZYYI $ \\ \hline     \end{tabular}     \end{table}     

 \begin{table} \caption{Qubit-wise commuting decomposition of the \ce{N2} molecular Hamiltonian at a separation of $ 1.87\angstrom$} \label{H8}      \begin{tabular}{|p{0.06\linewidth} | p{0.9\linewidth}|}     \hline     \textbf{Clique Index} & \textbf{QWC Hamiltonian Terms} \\     \hline   
 Identity & $-106.22020 \cdot IIIII$ \\ \hline
 0 & $ - 0.30710 \cdot IIIIZ + 0.00359 \cdot IIIZI + 0.01437 \cdot IIIZZ + 0.00359 \cdot IIZII + 0.01437 \cdot IIZIZ + 0.51603 \cdot IIZZI + 0.00359 \cdot IZIII + 0.01437 \cdot IZIIZ + 0.45287 \cdot IZIZI + 0.47393 \cdot IZZII + 0.00359 \cdot IZZZI + 0.01437 \cdot IZZZZ + 0.04515 \cdot ZIIII + 0.01587 \cdot ZIIIZ + 0.45323 \cdot ZIIZI + 0.47343 \cdot ZIZII + 0.04515 \cdot ZIZZI + 0.01587 \cdot ZIZZZ + 0.26725 \cdot ZZIII + 0.04515 \cdot ZZIZI + 0.01587 \cdot ZZIZZ + 0.04515 \cdot ZZZII + 0.01587 \cdot ZZZIZ + 0.51533 \cdot ZZZZI $ \\ \hline1 & $ - 0.20771 \cdot IIIIX - 0.12404 \cdot IIYYI - 0.01010 \cdot IYYII + 0.01010 \cdot XIYYI + 0.09373 \cdot XYIYI - 0.10383 \cdot XYYII $ \\ \hline2 & $ + 0.10383 \cdot IZZXI + 0.01016 \cdot XZZII + 0.01010 \cdot XZZXI $ \\ \hline3 & $ - 0.01010 \cdot IXZZI - 0.01016 \cdot XIZZI + 0.12404 \cdot XXZZI $ \\ \hline4 & $ - 0.10383 \cdot ZIZXI + 0.01010 \cdot ZXZII - 0.01047 \cdot ZXZXI $ \\ \hline5 & $ + 0.12404 \cdot ZZYYI $ \\ \hline     \end{tabular}     \end{table}     

 \begin{table} \caption{Qubit-wise commuting decomposition of the \ce{N2} molecular Hamiltonian at a separation of $ 2.00\angstrom$} \label{H9}      \begin{tabular}{|p{0.06\linewidth} | p{0.9\linewidth}|}     \hline     \textbf{Clique Index} & \textbf{QWC Hamiltonian Terms} \\     \hline    
 Identity & $-106.87553 \cdot IIIII$ \\ \hline
 0 & $ - 0.12236 \cdot IIIIZ - 0.12236 \cdot IIIZI + 0.11832 \cdot IIIZZ - 0.03042 \cdot IIZII - 0.01196 \cdot IIZIZ + 0.01092 \cdot IIZZI - 0.03042 \cdot IZIII + 0.01092 \cdot IZIIZ - 0.01196 \cdot IZIZI + 0.12814 \cdot IZZII + 0.12814 \cdot IZZZZ - 0.03042 \cdot ZIIII - 0.01196 \cdot ZIIIZ + 0.01092 \cdot ZIIZI - 0.01031 \cdot ZIZII - 0.01031 \cdot ZIZZZ + 0.01029 \cdot ZZIII + 0.01029 \cdot ZZIZZ - 0.01196 \cdot ZZZIZ + 0.01092 \cdot ZZZZI - 0.03042 \cdot ZZZZZ $ \\ \hline1 & $ - 0.00854 \cdot IIIIX + 0.00955 \cdot IIIXI + 0.11164 \cdot IIIXX + 0.01015 \cdot IIXII + 0.00955 \cdot IIXIX - 0.00854 \cdot IIXXI + 0.01042 \cdot IXIII + 0.00854 \cdot IXIIX - 0.00955 \cdot IXIXI - 0.12794 \cdot IXXII + 0.12794 \cdot XIIII + 0.00955 \cdot XIIIX - 0.00854 \cdot XIIXI - 0.01042 \cdot XIXII - 0.01015 \cdot XXIII $ \\ \hline2 & $ - 0.11164 \cdot IIIYY + 0.00955 \cdot IIYIY + 0.00854 \cdot IIYYI - 0.00854 \cdot IYIIY - 0.00955 \cdot IYIYI + 0.12794 \cdot IYYII + 0.00955 \cdot YIIIY + 0.00854 \cdot YIIYI - 0.01042 \cdot YIYII + 0.01015 \cdot YYIII $ \\ \hline3 & $ + 0.12794 \cdot XZZZZ $ \\ \hline     \end{tabular}     \end{table}

\end{document}